\documentclass[cernpreprint, atlasdraft=false, UKenglish, texmf, orcidlogo]{atlasdoc}
 
\usepackage{atlaspackage}
\usepackage{atlasbiblatex}
 
\usepackage{atlasphysics}

\graphicspath{{logos/}{figures/}}
 
\usepackage{ANA-STDM-2018-36-PAPER-defs}

\AtlasTitle{Measurement of the cross-sections of the electroweak and total production of a $Z \gamma$ pair in association with two jets
in $pp$ collisions at {\mbox{$\sqrt{s}$  = 13 TeV}} with the ATLAS detector}

\AtlasAbstract{
This Letter presents the measurement of the fiducial and differential cross-sections of the electroweak production of a $Z \gamma$ pair in association with two jets.
The analysis uses 140 fb$^{-1}$ of LHC proton--proton collision data taken at $\sqrt{s}$=13 TeV recorded by the ATLAS detector during the years 2015--2018.
Events with a $Z$ boson candidate decaying into either an  $e^+e^-$ or $\mu^+ \mu^-$ pair, a photon and two
jets  are selected.
The electroweak component is extracted by  requiring
a large dijet invariant mass and by using the information about the centrality of the system and is measured with an observed and expected significance well above five standard deviations.
The fiducial  $pp \rightarrow Z \gamma jj$ cross-section  for the electroweak production is measured to be 3.6 $\pm$ 0.5~fb.
The total  fiducial cross-section that also includes contributions where
the jets arise from strong interactions is measured to be $16.8^{+2.0}_{-1.8}$~fb.
The results are consistent with the Standard Model predictions. Differential cross-sections are also measured using the same events and are compared with parton-shower Monte Carlo simulations. Good agreement is observed between data and predictions.
}
 
\author{The ATLAS Collaboration}
 
\AtlasRefCode{ANA-STDM-2018-36}
 
\PreprintIdNumber{CERN-EP-2023-098}

\AtlasJournalRef{\PLB 846 (2023) 138222}
\AtlasDOI{10.1016/j.physletb.2023.138222}

\hypersetup{pdftitle={ATLAS document},pdfauthor={The ATLAS Collaboration}}
 
\begin{document}
 
\maketitle
 
\tableofcontents

% The next lines are included from the .//texfiles/1.Introduction.tex input file
\section{Introduction}
\label{sec:Introduction}
 
The study of the electroweak (EW) production of two vector bosons associated with two jets is a powerful test of the Standard Model (SM) due to its sensitivity to the gauge-boson self-interactions, related to the non-Abelian structure of the electroweak interaction.
It provides the means to investigate vector-boson scattering (VBS) processes ($VV\rightarrow VV$ with $V=W$, $Z$ or $\gamma$),
which directly probe the electroweak symmetry breaking sector of the SM~\cite{PhysRevLett.38.883}, and to extract constraints on anomalous
gauge-boson couplings~\cite{PhysRevD.74.073005}.
Improved constraints probe scales of new physics in the multi-\TeV\ range and provide
a way to look for signals of new physics in  a model-independent way.
 
\begin{figure}
\centering
\subfloat[]{\includegraphics[width=0.3\columnwidth]{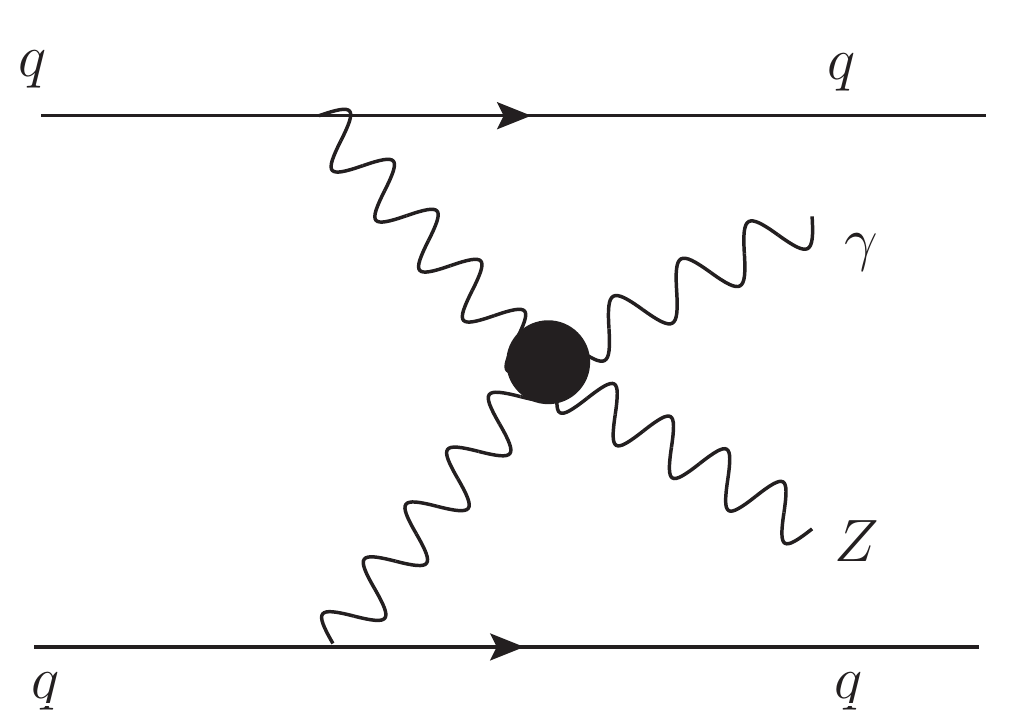}}
\subfloat[]{\includegraphics[width=0.3\columnwidth]{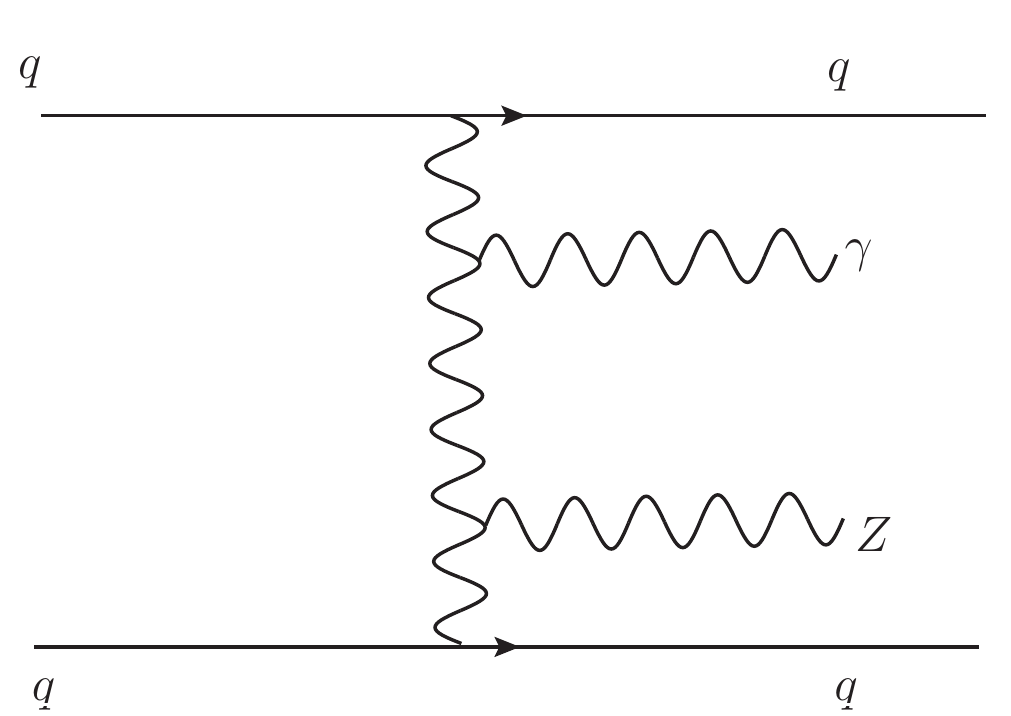}}
\subfloat[]{\includegraphics[width=0.3\columnwidth]{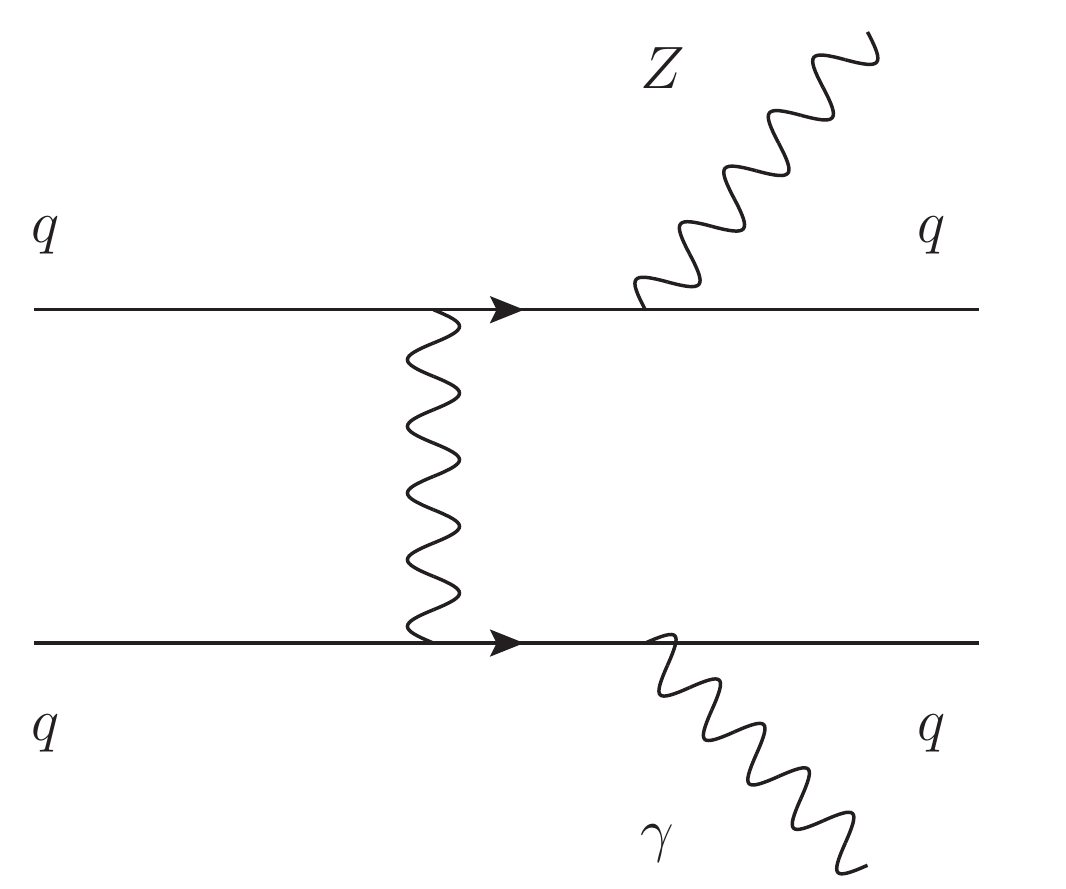}}\\
\subfloat[]{\includegraphics[width=0.3\columnwidth]{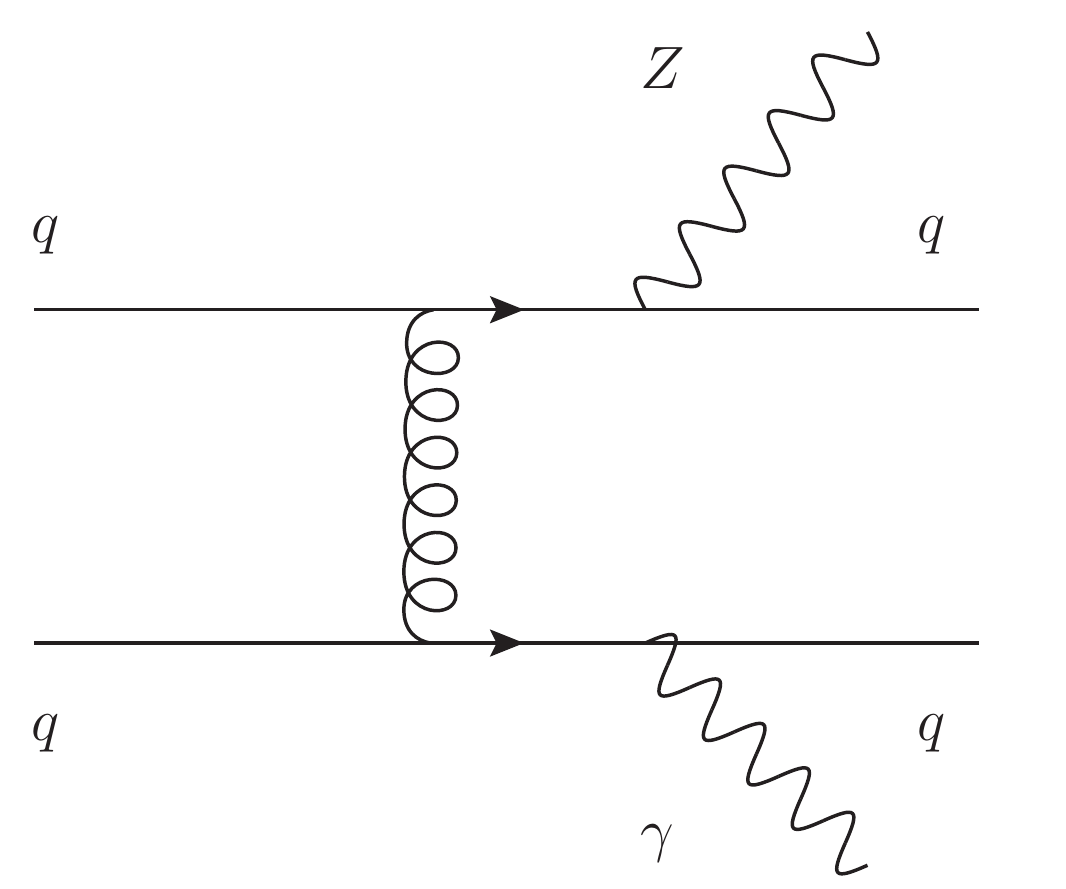}}
\subfloat[]{\includegraphics[width=0.3\columnwidth]{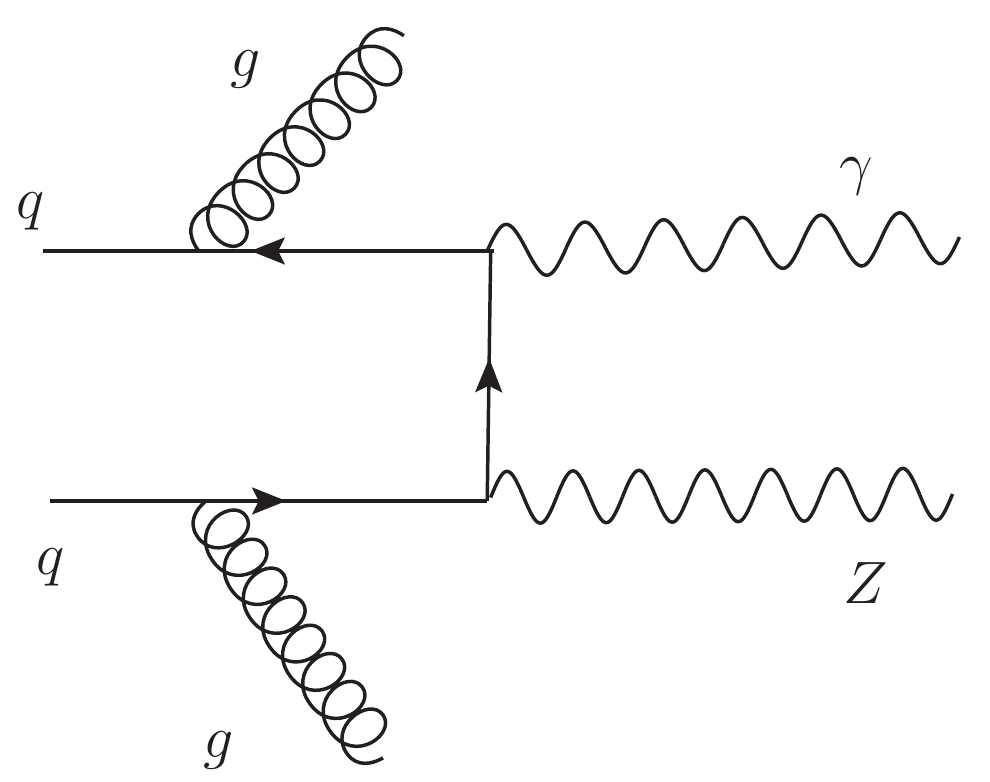}}
\caption{Representative Feynman diagrams of the processes relevant to this analysis: (a) quartic gauge coupling VBS, (b) triple gauge coupling VBS, (c) electroweak non-VBS, QCD-induced process with (d) gluon exchange or (e) gluon radiation.}
\label{fig:fey_diag}
\end{figure}

In particular, the study of the EW production of a $Z \gamma$ pair associated with two jets (referred to as $\textit{EW-Z}\gamma jj$)  is
interesting because it  probes the neutral quartic gauge couplings, as for the $ZZ$
production but with a larger expected cross-section.
These couplings are forbidden at the lowest order in the SM\@.
The EW production of the $Z\gamma jj$ final state, shown in the top row of Figure~\ref{fig:fey_diag}, consists of both VBS processes
directly  sensitive to triple and quartic gauge couplings,
and non-VBS processes, which incorporate
other EW contributions.

In the VBS process two jets are typically present, one in the forward direction and the other in the  backward direction, while the vector-boson pair is more centrally
produced~\cite{Rainwater:1996ud}. For such events the scattered quarks are not colour connected and little hadronic activity is expected in the gap between the two jets.
This topology allows VBS production to be distinguished statistically from the production of $Z\gamma{}jj$ final states via
mixed EW and quantum chromodynamics (QCD) mechanisms, referred to as $\textit{QCD-Z}\gamma jj$.
The bottom row of Figure~\ref{fig:fey_diag} shows examples of some $\textit{QCD-Z}\gamma jj$ diagrams where the strong interaction acts between the initial quarks, or where the jets arise from the strong interaction.
The $Z\gamma jj$ production via EW and QCD mechanisms interfere constructively when the initial and final states are the same, with an interference term at the level of 7\%.

Previous experimental results of
$\textit{EW-Z}\gamma jj$  production with the $Z$ decaying into charged leptons were published by the ATLAS and CMS collaborations using data collected at
$\sqrt{s}$ = 8~\TeV~\cite{STDM-2015-21, CMS-SMP-14-018}. Evidence of the process was reported by both experiments using partial data sets collected
at  $\sqrt{s}$ = 13~\TeV\ using 36~fb$^{-1}$~\cite{STDM-2017-26,CMS-SMP-18-007}, and the process has been observed by  CMS~\cite{CMS:2021gme} using the full Run 2 data sample. The measurement of the total $Z\gamma jj$ cross-section has been reported recently by ATLAS~\cite{ATLAS:2022wnf}. The results presented here complements this previous paper by providing a measurement of the total $Z\gamma jj$ in a VBS-like region, which is useful to obtain a detailed characterization of this process in a region sensitive to new physics.
 
The analysis described here exploits the full data sample collected with the ATLAS detector
in Run~2 at a centre-of-mass energy of  $\sqrt{s}$ =  13~\TeV\ corresponding
to an integrated luminosity of 140~fb$^{-1}$.
This Letter reports the observation by ATLAS of the $\textit{EW-Z}\gamma jj$ process, where the $Z$ boson decays into either $e^+e^-$ or $\mu^+ \mu^- $ pairs, and its fiducial and differential cross-section measurements in several observables:  the transverse momentum of the leading lepton ($p_{\mathrm{T}}^{l}$, sensitive to process modelling), the transverse momentum of the jets ($p_{\mathrm{T}}^{j}$), the invariant mass of and absolute rapidity difference between the two leading jets ($m_{jj}$ and $|\Delta y|$, sensitive to the $\textit{EW-Z}\gamma jj$  and $\textit{QCD-Z}\gamma jj$ kinematic differences), the transverse momentum of the photon and $Z\gamma$ systems and the absolute azimuthal difference between the $Z\gamma$ system and the two leading jets ($E_{\mathrm{T}}^{\gamma}$, $p_{\mathrm{T}}^{Z\gamma}$ and $|\Delta \phi(Z\gamma, jj)|$, potentially sensitive to new physics effects).
 
The fiducial and differential cross-sections that include the $\textit{QCD-Z}\gamma jj$ contribution are also reported for the same observables, in addition to the $Z$ boson transverse momentum ($p_{\mathrm{T}}^{\mathrm{Z}}$) and the centrality of the $Z\gamma jj$ system ($\zeta(Z\gamma$), described in Section~\ref{sec:Reconstruction}).

The measurement of the EW and total fiducial cross-sections presented in this Letter improves upon the precision of the previous ATLAS result~\cite{STDM-2017-26} and several variables are measured for the first time differentially in these processes ($p_{\mathrm{T}}^{Z\gamma}$, $|\Delta \phi(Z\gamma, jj)|$, $p_{\mathrm{T}}^{\mathrm{Z}}$ and $\zeta(Z\gamma$).

The layout of the Letter is as follows: the ATLAS detector is briefly described in Section~\ref{sec:ATLASDet},
the data sample and the simulated signal and background Monte Carlo (MC) samples used in the analysis are presented in Section~\ref{sec:DataMCSamples},
while the event reconstruction and selection are reported in Section~\ref{sec:Reconstruction}.
The determination of the background and event yields are discussed in Section~\ref{sec:Backgrounds}
and the experimental and theoretical uncertainties are presented in Section~\ref{sec:Systematics}.
The procedure
to extract the signal and to measure the differential cross-sections are described in Sections~\ref{sec:Procedure} and~\ref{sec:PhaseSpace}.
Finally, Section~\ref{sec:Results} presents the cross-section measurements, and conclusions are drawn in Section~\ref{sec:Conclusions}.

\newcommand{\AtlasCoordFootnote}{
ATLAS uses a right-handed coordinate system with its origin at the nominal interaction point (IP)
in the centre of the detector and the $z$-axis along the beam pipe.
The $x$-axis points from the IP to the centre of the LHC ring,
and the $y$-axis points upwards.
Cylindrical coordinates $(r,\phi)$ are used in the transverse plane,
$\phi$ being the azimuthal angle around the $z$-axis.
The pseudorapidity is defined in terms of the polar angle $\theta$ as $\eta = -\ln \tan(\theta/2)$.
Angular distance is measured in units of $\Delta R \equiv \sqrt{(\Delta y)^{2} + (\Delta\phi)^{2}}$.
When dealing with massive jets and particles, the rapidity $y = \frac{1}{2} \ln \left( \frac{E+p_z}{E-p_z} \right)$
is used, where $E$ is the jet/particle energy and $p_z$ is the $z$-component of
the jet/particle momentum. }

% End of text imported from the .//texfiles/1.Introduction.tex input file

% The next lines are included from the .//texfiles/2.ATLASDet.tex input file
\section{ATLAS detector}
\label{sec:ATLASDet}
 
The ATLAS detector~\cite{PERF-2007-01} at the LHC is a multipurpose particle detector with a forward--backward symmetric cylindrical
geometry{\footnote{ATLAS uses a right-handed coordinate system with its origin at the nominal interaction point (IP)
in the centre of the detector and the $z$-axis along the beam direction.
The $x$-axis points from the IP to the centre of the LHC ring,
and the $y$-axis points upwards.
Cylindrical coordinates $(r,\phi)$ are used in the transverse ($x,y$) plane,
$\phi$ being the azimuthal angle around the $z$-axis.
The pseudorapidity is defined in terms of the polar angle $\theta$ as $\eta = -\ln \tan(\theta/2)$.
Angular distance is measured in units of $\Delta R \equiv \sqrt{(\Delta y)^{2} + (\Delta\phi)^{2}}$.
} }and nearly $4\pi$ coverage in  solid angle. It consists of an inner tracking detector (ID),
electromagnetic and hadronic calorimeters, and a muon spectrometer (MS).
The ID, surrounded by a thin superconducting solenoid delivering a 2~T magnetic field,
provides precision tracking of charged particles
and momentum measurements in the pseudorapidity range of $|\eta| < 2.5$.
A high-granularity electromagnetic (EM) sampling calorimeter covers the pseudorapidity
range of $|\eta| < 3.2$, and a coarser granularity calorimeter up to $|\eta| = 4.9$.
The hadronic calorimeter system covers the entire pseudorapidity range up to  $|\eta| = 4.9$.
The MS consists of three large superconducting toroids each containing eight coils, a system of trigger
chambers, and precision tracking chambers, which provide trigger and tracking capabilities in the
range $|\eta| < 2.4$ and $|\eta| < 2.7$, respectively.
A two-level trigger system~\cite{TRIG-2019-04} is used to select events. The first-level
trigger is implemented in hardware and uses a subset of the detector
information. This is followed by the software-based high-level trigger system,
which runs offline reconstruction.
 
An extensive software suite~\cite{ATL-SOFT-PUB-2021-001} is used in data simulation, in the reconstruction and analysis of real and simulated data, in detector operations, and in the trigger and data acquisition systems of the experiment.

% End of text imported from the .//texfiles/2.ATLASDet.tex input file

% The next lines are included from the .//texfiles/3.DataMCSamples.tex input file
\section{Simulated event samples}
\label{sec:DataMCSamples}
 
MC simulated samples are used to model the $\textit{EW-Z}\gamma jj$ signal and a variety of background processes.
The signal process was generated at leading-order (LO) accuracy (at order $\alpha_{EW}^4$, where $\alpha_{EW}$ is the electroweak coupling constant)
using \MGNLO[2.6.5]~\cite{Alwall:2014hca} with the default dynamical scale choice
and the NNPDF3.1 LO
parton distribution function (PDF) set~\cite{Ball:2014uwa}.
\PYTHIA 8.240~\cite{Sjostrand:2014zea} with the `dipoleRecoil' option turned on, and configured with the A14 set of tuned parameters (tune)~\cite{ATL-PHYS-PUB-2014-021}, was used to add parton-showering, hadronisation and underlying event activity.
The signal MC was also interfaced with \Herwig++ 2.7.1~\cite{Bahr:2008pv, Herwig:27} for parton showering, hadronisation and underlying event activity. The comparison between the two samples is used to
estimate the uncertainties due to the choice of the parton showering and underlying event models.
 
The dominant background in the cross-section measurement of the EW production of  $Z \gamma jj$ events is represented by the $\textit{QCD-Z}\gamma jj$ process.
Two sets of MC samples are used
to model this final state.
The nominal sample was produced with \SHERPA 2.2.11~\cite{Sherpa2211, Bothmann_2019}, where matrix elements were calculated with up to one additional parton at next-to-leading order (NLO) and up to three additional partons at LO\@. The matrix element calculation included all diagrams at order $\alpha_{EW}^2$. The virtual QCD corrections for matrix elements at NLO accuracy were provided by
the \OPENLOOPS\ library~\cite{Buccioni:2019sur}.  The merging of the matrix element and parton shower (PS) was performed with MEPS@NLO~\cite{Hoeche:2012yf, Catani:2001cc}. The NNPDF3.0 next-to-next-to-leading-order (NNLO) PDF was used in conjunction with a dedicated PS tuning developed by the \SHERPA authors.
An alternative sample was produced
with  \MGNLO[2.3.3] and used for cross checks. This sample has the default dynamical scale using the
NNPDF3.0 NLO PDF set.
The matrix element calculation in this sample includes
all diagrams   at order $\alpha_{EW}^2$
and the emission of up to two extra final-state partons, where up to one additional final-state parton is at NLO.
 
Additionally, for the evaluation of the theoretical uncertainty, a set of four samples was generated at particle level using \SHERPA 2.2.11 with the NNPDF3.0 NNLO PDF set, as well all other generation parameters being the same as the nominal \SHERPA 2.2.11 sample,  in order to provide
results with alternative merging and resummation scales. Two samples were produced with merging scale variations (QCUT=15~\GeV\ and QCUT=30~\GeV) and two samples with resummation scale factors (QSF=0.25 and QSF=4).\footnote{ QCUT indicates the scale  for the calculation of the overlap between jets from the matrix element and the parton shower and is nominally 20~\GeV, and
QSF represents the scale used for the resummation of the soft gluon emissions and is nominally 1.}

Interference between the EW and QCD processes was estimated at LO accuracy using the \MGNLO[2.3.3] MC event generator with the NNPDF3.0 LO PDF set including contributions
to the sum of the amplitudes of the matrix element squared
at order $\alpha_{S}\alpha_{EW}^3$. These interference effects are found to be positive and about 7\% of the $\textit{EW-Z}\gamma jj$ cross-section in the fiducial phase space studied. This effect is included as a systematic uncertainty in the signal prediction.
 
In all samples described above, photon isolation criteria were imposed at parton level making use of the smooth-cone isolation prescription introduced in Ref.~\cite{Frixione:1998jh}.
This procedure removes contributions in which the photon is produced from quark or gluon fragmentation in an infrared safe way to all orders of perturbation theory.
The chosen isolation parameters are $\delta_0=0.1$, $\epsilon=0.1$ and $n=2$.
 
The second-largest background, arising from the $Z$+jets process with one of the jets misidentified as a photon, is estimated with a data-driven method.  A MC sample is only used to estimate a correlation factor between different control regions as explained in Section~\ref{sec:Backgrounds}.
This sample was produced with \POWHEGBOX v1~\cite{powheg1, powheg2, powheg3} at NLO accuracy with the CT10~\cite{Guzzi:2011sv} NLO PDF set, interfaced to \PYTHIA 8.210~\cite{Sjostrand:2014zea} with the AZNLO tune~\cite{STDM-2012-23}.
 
The third-largest background, arising from the $t\bar{t}\gamma$ process was generated at LO accuracy with \MGNLO using the NNPDF 2.3 LO PDF set~\cite{Ball:2012cx} and was interfaced to the \PYTHIA 8.212 generator, configured with the A14 tune~\cite{ATL-PHYS-PUB-2014-021}.
An NLO factor of 1.44 was applied, based on the value found in an analysis of $t\bar{t}\gamma$ production at $\sqrt{s}=13$~\TeV\ by the ATLAS Collaboration~\cite{TOPQ-2017-14}, which normalizes the LO prediction from this MC sample to an NLO calculation in the fiducial phase-space region used in the $t\bar{t}\gamma$ analysis in the dilepton channel.
The predicted contribution from this background is validated using an $e\mu\gamma$ control region as explained in Section~\ref{sec:Backgrounds}.
 
All other backgrounds are smaller and are estimated with MC simulation.
The background process $\textit{QCD-WZ}(l\nu ll)$ was generated with \SHERPA 2.2.2 at NLO with up to one additional parton, using the NNPDF3.0 NNLO PDF set.
The $\textit{EW-WZ}(l\nu ll)jj$ background was generated with \MGNLO[2.6.2] at LO accuracy, using the NNPDF3.0 LO PDF set and  interfaced to \PYTHIA 8.235.
The $WW\gamma$ background is only considered in the $e\mu\gamma$ control region study and was generated with \SHERPA 2.2.5 at NLO, using the NNPDF3.0 NNLO PDF set.

The simulated samples are overlaid with additional proton$-$proton interactions (pile-up) generated with \PYTHIA 8.186 using
the A3 tune~\cite{ATL-PHYS-PUB-2016-017} and the NNPDF2.3LO PDF set~\cite{Ball:2012cx}.
MC events are  reweighted to better reproduce the distribution of the mean number of interactions per bunch crossing
observed in data.
All generated events were passed through the ATLAS detector simulation~\cite{SOFT-2010-01} based on \GEANT~\cite{Agostinelli:2002hh} and processed using the same reconstruction software
as for data.
Scale factors are applied to the simulated events to correct for small differences between them and data in the trigger, reconstruction, identification and isolation efficiencies
for photons, electrons and muons. Furthermore, in simulated events  electron, photon and jet energy and the muon momentum are smeared to account for the small
differences in resolution between data and simulation.

% End of text imported from the .//texfiles/3.DataMCSamples.tex input file

% The next lines are included from the .//texfiles/4.Reconstruction.tex input file
\section{Event reconstruction and selection}
\label{sec:Reconstruction}
 
The data were collected between 2015 and 2018 during proton--proton collisions at $\sqrt{s}$ = 13~\TeV.
The integrated luminosity of the sample used for the analysis is 140 fb$^{-1}$. The sample only includes data recorded with stable beam conditions and with all relevant subdetector systems
operational~\cite{DAPR-2018-01}.
 
Events were selected using unprescaled single lepton and dilepton triggers~\cite{TRIG-2018-05, TRIG-2018-01} with transverse momentum ($p_{\mathrm{T}}$) thresholds
that depended on the lepton flavour and running period.
In 2015  a single-electron or muon trigger, with $p_{\mathrm{T}}$ above 24 and 20~\GeV\ respectively, was required
while in the following years these thresholds were set to 26~\GeV\  for both flavours of leptons.
Additional single-lepton triggers with higher $p_{\mathrm{T}}$ thresholds but with looser  identification criteria were also used
to increase the total data-taking efficiency. Events with a pair of electron candidates with  $p_{\mathrm{T}} > 12$~\GeV, or a pair
of muon candidates  satisfying $p_{\mathrm{T}} > 18$~\GeV\ and $p_{\mathrm{T}} > 8$~\GeV\ for the leading and subleading muons,
were also selected at trigger level in 2015. In the following years these dilepton trigger thresholds were increased  up to 24~\GeV\ for
the dielectron case and 22 (8)~\GeV\ for the leading (subleading) muon for the dimuon case.
The trigger efficiency for events satisfying all the selection criteria described below is about 99\%.
 
Events are required to have at least one collision vertex reconstructed from at least two tracks, where the tracks must have a
$p_{\mathrm{T}}$ larger than 500~\MeV. The hard-interaction vertex of the event is chosen as the one with the largest value of the sum
of  the squared transverse momentum of the associated tracks.
 
Electron candidates, reconstructed from topological clusters of energy deposited in the EM calorimeter that are matched to an ID track,
are required to satisfy the medium likelihood identification criterion of Ref.~\cite{EGAM-2018-01}.
This is based on a combination of shower shape information from the EM calorimeter
and tracking information from the ID.
Electron candidates are required to have $p_{\mathrm{T}} > 20$~\GeV\ and $|\eta| < 2.47$ but excluding the
transition region between the barrel and endcap electromagnetic calorimeters ($1.37 < |\eta| < 1.52$).
The overall efficiency of the electron reconstruction and identification is about 80$\%$  for electrons with  $p_{\mathrm{T}} \approx 20$~\GeV\ and
increases with $p_{\mathrm{T}}$.
 
Muon candidates, reconstructed by matching tracks in the ID with tracks in the MS, are required to satisfy the medium identification criterion of Ref.~\cite{MUON-2018-03}.
This includes requirements on the number of hits matched to the tracks reconstructed in the ID and in the MS, and on
the probability that the ID and MS momentum measurements are compatible.
Muon candidates are required to have $p_{\mathrm{T}} > 20$~\GeV\ and $|\eta| < 2.5$.
The overall efficiency of the muon reconstruction and identification is above 97$\%$
with no strong dependence on $p_{\mathrm{T}}$.
 
Electron and muon candidates are required to originate from the primary vertex. The significance of the transverse impact parameter,
defined as the absolute value of the track transverse impact parameter, $|d_0|$, measured relative to the hard-interaction vertex and divided by its uncertainty,
is required to be less than five for electrons and  less than three for muons.
Furthermore, for both electrons and muons the difference $\Delta z_0$ between the value of the $z$ coordinate of the point on the track at which $d_0$ is defined, and the $z$ position
of the primary vertex, is required to satisfy $| \Delta z_0 \cdot \sin \theta|  < 0.5$ mm (where $\theta$ is the track polar angle).
 
Photon candidates are reconstructed and identified using algorithms based on the expected shapes of showers developing in the electromagnetic calorimeter~\cite{EGAM-2018-01}
.
Both converted and unconverted candidates{\footnote{For the converted candidates, the photon cluster is matched to a reconstructed conversion vertex formed either from two oppositely
charged-particle tracks or from a single track consistent with having originated from a photon conversion. For the unconverted photons the photon cluster is matched to neither a conversion vertex
nor an electron track.} }are retained.
Photon candidates are selected
if they are reconstructed within the fiducial volume of the central calorimeter ($|\eta| < 2.37$) and  outside the transition region between the barrel and endcap electromagnetic calorimeters ($1.37 < |\eta| < 1.52$).
 
Photon, electron and muon candidates are required to be isolated from other particles.
In all cases, the isolation criteria are based on the sum, $p_{\mathrm{T}}^{\textit{iso}}$, of the scalar transverse momenta of tracks with $p_{\mathrm{T}} > 1$~\GeV, and on the sum,
$E_{\mathrm{T}}^{\textit{iso}}$, of the transverse energy of topological clusters, within cones of size $\Delta R$ around the photon or lepton candidates, excluding the contribution of the candidates themselves.
The calorimeter isolation is also corrected on an event-by-event basis  for the contribution from the underlying event and pile-up.
Electron candidates are required to satisfy the {\textit{FCLoose}} isolation criteria of Ref.~\cite{EGAM-2018-01} with a cone of size $\Delta R$ = 0.2. The efficiency of the isolation criteria is greater than 95\% for electrons with $p_{\mathrm{T}}>20$~GeV.
Muon candidates are required to satisfy the  {\textit{PflowLoose\_FixedRad}} isolation criteria of Ref.~\cite{MUON-2018-03} with a cone of size $\Delta R$ = 0.2. The efficiency of the isolation criteria is greater than 90\% for muons with $p_{\mathrm{T}}>20$~GeV.
 
Photon candidates are required to satisfy the {\textit{FixedCutLoose}} isolation criteria of Ref.~\cite{EGAM-2018-01}. The photon isolation criterion employs a cone of size $\Delta R$ = 0.2 for both the track and calorimeter isolation, and requires $p_{\mathrm{T}}^{iso}$/E$_{\mathrm{T}}^\gamma < 0.05$ and $E_{\mathrm{T}}^{iso}$/E$_{\mathrm{T}}^\gamma< 0.065$,
where E$_{\mathrm{T}}^\gamma$ is the photon transverse energy.
 
At least one isolated photon, satisfying tight identification requirements is required.  The efficiency of  the  tight photon identification criterion, for isolated photons, ranges from 80$-$85$\%$ for photons of transverse energy E$_{\mathrm{T}}^\gamma \approx 25$~\GeV\
depending on the pseudorapidity region of the  detector and on the conversion status of the candidate.
 
Jets are clustered using the anti-$k_t$ algorithm~\cite{Cacciari:2008gp, FASTJets} with a radius parameter of $R = 0.4$.
The inputs to the algorithm are obtained with a particle flow procedure using topological clusters in the calorimeter and reconstructed tracks~\cite{PERF-2015-09}.
 
Jets are calibrated and corrected for detector effects using a combination of simulated events and in situ methods.
Jet candidates are required to have $p_{\mathrm{T}} > 25$~\GeV\ and rapidity $|y| < 4.4$.
Jets with $p_{\mathrm{T}} < 60$~\GeV\ and $|\eta| < 2.4$ are required to be consistent with originating from the primary vertex
using the tight working point of the jet vertex tagging algorithm of Ref.~\cite{PERF-2014-03}.
 
A procedure to remove ambiguities in the particle reconstruction is applied:
jet candidates are removed if they overlap with electron or photon candidates, i.e. $\Delta R (j, e) < 0.2$ or $\Delta R (j, \gamma) < 0.4$, then leptons are removed if they are close to a jet candidate, i.e. $\Delta R (\ell, j) < 0.4$ ($\ell$ = e, $\mu$),
photons are removed if  they are close to a lepton candidate, i.e. $\Delta R (\gamma, \ell) < 0.4$ and finally electron candidates are removed if they overlap with muon candidates i.e. $\Delta R (\mu, e) < 0.2$.

Events are required to have exactly two leptons of same flavour and opposite charge, at least one photon
and at least two jets.
One of the electrons or muons in the lepton pair must be matched to the electron or muon that triggered the event.
Events are further selected by requiring that the leading lepton has $p_{\mathrm{T}} > 30$~\GeV\ and that the leading photon has $p_{\mathrm{T}} > 25$~\GeV\ and satisfies isolation and tight identification requirements.  To remove
contributions from low-mass resonances,  the invariant mass $m(\ell \ell)$ of the opposite-charge, same-flavour lepton pair must be larger than 40~\GeV.

To suppress events  originating from leptonic $Z$ decays where one of the leptons has radiated a photon,
the sum of $m_{\ell \ell}$ and the invariant mass  of the $\ell^+\ell^- \gamma$ system, $m_{\ell^+\ell^- \gamma}$,  formed from the lepton pair and the highest-$E_T^\gamma$ photon candidate,
must be larger than 182~\GeV, approximately twice the mass of the Z boson, as adopted in previous publications~\cite{STDM-2015-21, STDM-2017-26}.
 
Furthermore, to enhance the VBS topology, events must have at least two jets with $p_{\mathrm{T}}^{j}$ above 50~\GeV\ and a rapidity difference between them, $|\Delta y| > 1$.
The invariant mass of this pair of jets, $m_{jj}$, is required to be larger than 150~\GeV\ for the total $Z\gamma jj$ process measurements, and larger than 500~\GeV\ for the $Z\gamma jj$ EW process measurements. This selection significantly reduces the number of events with three bosons in the final state in first case, and the number of QCD $Z\gamma jj$ background events in the second case.
 
Events containing $b$-tagged jets are rejected.
The $b$-tagging algorithm provides a working  point with a 70$\%$ selection efficiency for $b$-jets in an inclusive
$t\bar{t}$ MC sample and rejection factors of $\approx$ 10 and 400 for charm- and light-flavour jets, respectively~\cite{FTAG-2018-01}.
The two highest-\pT jets satisfying these conditions are referred to as VBS tagged jets.
Events with additional jets of transverse momentum above 25~\GeV\ in the rapidity gap between the two VBS tagged jets  are rejected.
The centrality of the $\ell^+\ell^- \gamma$ system relative to the VBS tagged jets ($j_1$ and $j_2$) defined as
\begin{equation}
\zeta(Z\gamma) = \left| \frac{y_{Z\gamma}-(y_{j_1}+y_{j_2})/2}{y_{j_1}-y_{j_2}} \right|,
\end{equation}
where $y$ indicates the rapidity,  is required to be less than 5. \\
 
For the EW $Z\gamma jj$ signal extraction, within the $m_{jj}>500$~GeV region,  the selected events are further split into a signal region (SR, $\zeta(Z\gamma) < 0.4$)
and a QCD control region (CR, $\zeta(Z\gamma) > 0.4$)
as explained in Section~\ref{sec:Procedure}. For the measurements of the full $Z\gamma jj$ process, within the relaxed $m_{jj}>150$~GeV region, only the region $\zeta(Z\gamma) < 0.4$ is used, referred to as `Extended SR'.
This variable has been chosen to build the signal and control regions because it has been found to be almost uncorrelated with $m_{jj}$.

The observed total number of events in the $m_{jj}>500$~\GeV\ SR and CR is 562 and 274 respectively. In the $m_{jj}>150$~\GeV\ Extended SR phase space, the observed total number of events is 1461.

% End of text imported from the .//texfiles/4.Reconstruction.tex input file

% The next lines are included from the .//texfiles/6.Background.tex input file
\section{Background estimation}
\label{sec:Backgrounds}
 
The main source of background in the  cross-section measurement of the EW production of
$Z \gamma jj$ final states  consists of $Z \gamma jj$ events  from  QCD-induced processes.
The shape of this background is estimated from simulation and
the normalisation is determined simultaneously with the signal strength via a maximum-likelihood fit
to the $m_{jj}$ data distribution in the SR and CR that are defined in Section~\ref{sec:Reconstruction}.
A $\textit{QCD-Z}\gamma jj$ normalisation parameter, together with the signal normalisation
is extracted and the CR is used  to constrain the systematic uncertainties in both $\textit{QCD-}Z \gamma jj$ and  $\textit{EW-Z}\gamma jj$ processes.
The fit procedure is described in Section~\ref{sec:Procedure}.
 
The second-largest background (and largest background for the total $Z\gamma jj$ cross-section measurements) arises from the $Z$+jets process with a jet misidentified as a photon and is referred to as non-prompt photon background.
This contribution is estimated in data separately in the SR, Extended SR and CR using a two-dimensional sideband
method~\cite{STDM-2010-08} similar to that
applied in the previous analyses~\cite{STDM-2015-21, STDM-2017-26} and  includes the background deriving from both EW and QCD $Z$+jets induced processes.
In this procedure the selection criteria that define the SR, Extended SR and CR are applied to data
except for the photon identification and calorimeter isolation requirements.
Photon candidates are split into those that satisfy the tight ID requirements and those that do not.
The candidates that fail to satisfy the tight identification requirements are required to satisfy a non-tight selection criterion that removes requirements on four of the nine EM calorimeter shower shape variables required for tight photons.
These two samples are further split according to whether the photon satisfies the calorimeter isolation criteria or not.
A prompt photon region and three control regions are then defined using this method. The number of  $Z$+jets events in SR, CR and Extended SR is obtained from the number of events in the three control regions by assuming that the ratio of non-prompt isolated and non-isolated photon candidates is the same for tightly identified photons
and for photons failing to satisfy the tight identification criteria. The small  residual correlation between the two variables and
the leakage of prompt photon candidates into the non-prompt photon region are estimated from simulation.
The correlation is also estimated in data using a control region where the photon fails track isolation and the  difference between the MC and data results are included in
the systematic uncertainty (Section~\ref{sec:Systematics}).
The shape of this background is obtained from both control regions where
the photon candidate fails to satisfy the tight identification criteria.
Comparisons with different control regions show that
this choice
does not  introduce any bias to the shape of the distributions.
 
The third-largest background arises from the $t\bar{t}\gamma$ process. It is estimated from simulation and checked by comparing
predictions with data using an $e \mu \gamma$ data sample where the same selections are applied as those that define the SR, Extended SR and CR, except that a different-flavour lepton pair is chosen instead of a same-flavour pair.
The very small number of non-$t\bar{t}\gamma$ events in this sample is estimated either from simulated events, for events with a prompt photon, or
with the procedure described above
for events with a jet misidentified as a photon.
Predictions are compared with data in the control regions $e\mu\gamma$\_SR, $e\mu\gamma$\_Extended\_SR and $e\mu\gamma$\_CR, before or after requiring
that the events have at least one $b$-jet.
In both cases it is found that predictions must be scaled by a factor of 1.44$\pm$ 0.22 (i.e., an uncertainty of $\pm$ 15\%) to describe the data well, in agreement with previous studies~\cite{TOPQ-2017-14}.

The background contribution due to $WZjj$ events is minor and is evaluated from simulation while the contribution from  other processes is found to be negligible.
In the SR, it is estimated that 48\% of the events come from $\textit{EW-Z}\gamma jj$ and 44\% from $\textit{QCD-Z}\gamma jj$, compared to 9\% and 81\% for $\textit{EW-Z}\gamma jj$ and $\textit{QCD-Z}\gamma jj$ in the CR, respectively.
In the Extended SR, it is estimated that 88\% of events come from $Z\gamma jj$.
 
The yields of the different sources of background, after the fit to extract the signal is performed, are shown in Table~\ref{tab:BackgroundSummary}.

% End of text imported from the .//texfiles/6.Background.tex input file

% The next lines are included from the .//texfiles/7.Systematics.tex input file
\section{Systematic uncertainties}
\label{sec:Systematics}
The overall uncertainties in the differential cross-section measurements are dominated by the statistical uncertainty in the data, and the inclusive cross-section measurement uncertainties are shared equally between the statistical and systematic uncertainties.
 
Systematic uncertainties that affect the acceptance and the shape of the $m_{jj}$ distribution for the fiducial cross-section measurement and of the other observables for the differential cross-section measurements for both signal and backgrounds are considered.
The $\textit{EW-Z}\gamma jj$  and $\textit{QCD-Z}\gamma jj$  normalisations are extracted from a likelihood fit.
Systematic uncertainties in the shapes and normalisation of distributions are only considered if they are
found to make an impact on the result, which is translated into a threshold for only considering the uncertainties that are larger than 0.5\% for all measurements except for the $\textit{EW-Z}\gamma jj$ differential cross-section for which the threshold is 1\%, due to larger statistical uncertainties in this measurement. Uncertainties that are smaller than these thresholds are found to make no difference to the results when added to the fit.
 
The experimental systematic uncertainties that are accounted for in the analysis
include uncertainties in the energy scale
and resolution of jets, photons and electrons, in the scale and resolution of the muon momentum and uncertainties in the scale factors applied to simulation to reproduce the trigger, reconstruction, identification, and isolation efficiencies measured in data. Uncertainties due to the suppression of pile-up jets and to the $b$-jet veto are also considered.
 
The largest of these experimental uncertainties, in all cross-section measurements, are related to the jet energy calibration and response, and are at the level of 3\% in most of the measured bins, but can reach 7\% in highest bin of the $m_{jj}$ differential measurement. The dominant uncertainties associated with photons are due to the identification and isolation efficiencies~\cite{PERF-2017-02}, which are both about 1$\%$ with a negligible dependence on  $m_{jj}$, but can reach 3\% in the highest bin of $p_{T}^{\gamma}$ differential cross-section measurement.
Uncertainties in the lepton reconstruction, identification, isolation,  trigger efficiency,  energy/momentum scale and resolution are determined using $Z\rightarrow \ell \ell$ events~\cite{EGAM-2018-01, PERF-2017-01, MUON-2018-03}.
 
The dominant contribution comes from the electron identification efficiency, which is about 1$\%$ with a negligible dependence on  $m_{jj}$, but can reach about 4\% in the highest bin of the $p_{T}^{l}$ differential measurement.
The uncertainty associated with the pile-up modelling depends on $m_{jj}$ and is 2$\%$ on average.
The uncertainty in the combined 2015--2018 integrated luminosity is 0.83\%~\cite{ATLAS:2022hro}, obtained using van der Meer beam separation scans during dedicated running periods.
 
The overall uncertainty related to the background estimation is the second largest experimental uncertainty in the cross-section measurements, at the level of 1-2\% depending on the variable or process measured.
A 35\% uncertainty on the normalisation of the $Z$+jets background is used. It is extracted along with the data-driven procedure described in Section~\ref{sec:Backgrounds} and it accounts for the uncertainty in the number of events in the two-dimensional
sideband method used to estimate this background (statistical component). It also includes
the uncertainty related to the estimate of the correlation between the photon identification and
isolation requirements and to the leakage of the prompt photons into the non-prompt regions.
The statistical component dominates. The uncertainty derived from the evaluation of the shape of the $Z$+jets background in different observables is found to be negligible.
A 15$\%$ and a  20$\%$ yield uncertainty, derived from the data-driven normalisation correction (see Section~\ref{sec:Backgrounds}) and from QCD scales and PDF variations, is assigned to the estimate of the normalisation of the  $t\bar{t}\gamma$ and $W^{\pm}Z\rightarrow lll\nu$  backgrounds, respectively.
Other sources of background, including one due to the superposition of pile-up events, are found  to be negligible and  therefore the related uncertainties are neglected.

The main theoretical uncertainties that are considered in the analysis are related to the scale and PDF set choices in the MC
generation of the signal and the $\textit{QCD-Z}\gamma jj$ background.
The effect of missing higher orders is estimated by changing  the default values of the renormalisation and
factorisation scales, $\mu_R$ and $\mu_F$,  by a factor of 0.5 and 2.0  with the constraint that the ratio $0.5  \leq \mu_R/\mu_F \leq 2$.
The maximal change in the shape of distributions from these variations  is taken as the associated uncertainty.
This procedure is performed for the $\textit{EW-Z}\gamma jj$  and $\textit{QCD-Z}\gamma jj$ processes using the nominal \MGNLO and \SHERPA 2.2.11 samples, respectively.
The uncertainties  due to the choice of PDF in the shape of distributions  for the
$\textit{EW-Z}\gamma jj$ and $\textit{QCD-Z}\gamma jj$ processes are
evaluated using the eigenvalues of the PDF set following the PDF4LHC prescription~\cite{Butterworth:2015oua}.
To account for the uncertainties related to the modelling of $\textit{QCD-Z}\gamma jj$,
the impact of the merging and resummation scale  is derived using the five  \SHERPA 2.2.11 samples with different QCUT and QSF scales as described in Section~\ref{sec:DataMCSamples}.
Signal uncertainties due to the choice  of the parton showering and the underlying event model are estimated by interfacing the signal MC
to either \PYTHIA or \Herwig and the five up and down eigenvariations of the A14 tune.
The parton shower uncertainty from taking the difference between \Herwig and \PYTHIA is dominant and has a strong shape component. Parton shower and underlying event uncertainties
are uncorrelated. The underlying event uncertainty is obtained by taking the envelope of the maximum variations bin by bin.

The interference between the EW signal and the $\textit{QCD-Z}\gamma jj$ background is not included as part of the EW signal in the fit that extracts the
EW signal. Instead this contribution is directly computed with \MGNLO and  the size and shape of the interference contribution is taken as an extra template uncertainty on the signal. The overall size effect is 7\% in the phase space dedicated to the signal measurement and the shape varies depending on the variable studied, with typically larger interference effect where the $\textit{QCD-Z}\gamma jj$ background dominates.
Another source of uncertainty arises from the finite size of the MC (at the level of 1\%) and the data samples (ranging from 9\% to ~22\% depending on the measurement, variable and bin considered).

The implementation of these uncertainties in the various measurements performed are described in Sections~\ref{sec:Procedure} and ~\ref{sec:PhaseSpace} and their final impact on the measurements is discussed in Section~\ref{sec:Results}. Table~\ref{tab:breakSystData}  provides a breakdown of the uncertainties in the final measurement.

% End of text imported from the .//texfiles/7.Systematics.tex input file

% The next lines are included from the .//texfiles/8.SignalExtraction.tex input file
\section{Signal extraction procedure}
\label{sec:Procedure}
To extract the $\textit{EW-Z}\gamma jj$ cross-section, the signal strength parameter defined as
 
\begin{equation}
\label{eq:Strength}
\mu_{EW} = \sigma^{EW}_{meas}/ \sigma^{EW}_{exp}\,
\end{equation}
 
is introduced in both the SR and CR, and obtained with a maximum-likelihood fit simultaneously to the data $m_{jj}$ distributions in both regions using template MC distributions.
In Eq.~(\ref{eq:Strength})
the numerator indicates the measured $\textit{EW-Z}\gamma jj$ cross-section
and the denominator is the expected  $\textit{EW-Z}\gamma jj$ cross-section.
An unconstrained normalisation parameter is introduced in the SR and CR for the $\textit{QCD-Z}\gamma jj$ contribution and is extracted simultaneously with $\mu_{EW}$.
 
The significance of observing the $\textit{EW-Z}\gamma jj$ process is estimated by using a profile likelihood ratio of the background only hypothesis ($\mu_{EW} = 0$) and the best fit result ($\mu_{EW} = \hat{\mu}$)~\cite{Cowan:2010js}. The  $\textit{EW-Z}\gamma jj$ cross-section is obtained from the signal strength by multiplying it by the MC cross-section prediction in the SR region defined at particle level.
 
The extraction of the full $\textit{Z}\gamma jj$ cross-section is performed in a very similar manner, adding together the relative fractions of the $\textit{EW-Z}\gamma jj$ and $\textit{QCD-Z}\gamma jj$ $m_{jj}$ templates, as predicted from MC, and defining the signal strength $\mu_{Z\gamma}$ as parameter of interest of the fit.  No CR is used in this fit, and a wider region (with $m_{jj}>150$~\GeV, the Extended SR) is used.
The $\textit{Z}\gamma jj$ cross-section is obtained by multiplying the signal strength by the MC cross-section prediction in the Extended SR defined at particle level.

In these two measurements, the electron and muon channels are combined directly in the input histograms, by summing the two contributions, such that a single template is used.
Probability density functions are built for the $m_{jj}$ templates in the SR, Extended SR and CR  based on a Poisson distribution and are combined in
an extended likelihood.  Each  source of uncertainty is implemented in these functions as a  nuisance parameter of the likelihood fit with a Gaussian constraint, except for the MC statistical uncertainty that is implemented with a Poisson constraint.
 
The uncertainties from the $Z$+jets, $WZ$ and $t\bar{t}\gamma$ backgrounds are treated as correlated between regions. All systematic uncertainties except for the theoretical ones are correlated between processes and between the two regions.
The PDF and scale uncertainties are not correlated between processes, and for the $\textit{EW-Z}\gamma jj$ process they are also not correlated between regions. Choosing a different correlation scheme does not change the result. The normalisation part of PDF and scale uncertainties are subtracted to consider only acceptance effects on the signal. The interference, parton shower and underlying event uncertainties in the $\textit{EW-Z}\gamma jj$ contribution are also not correlated between regions. The merging and resummation scale uncertainties in the $\textit{QCD-Z}\gamma jj$ contribution are correlated between regions.

For the $\textit{EW-Z}\gamma jj$ measurement, the difference in the predicted  $m_{jj}$ shape between two different $\textit{QCD-Z}\gamma jj$ MCs is the same in the SR and CR within modelling uncertainties;
therefore,  the $m_{jj}$ shape in the CR is used to validate the $m_{jj}$ shape of the $\textit{QCD-Z}\gamma jj$ background  in the SR, and to constrain the correlated systematic uncertainties.

\begin{figure}[h!]
\centering
\subfloat[]{\includegraphics[width=.35\columnwidth]{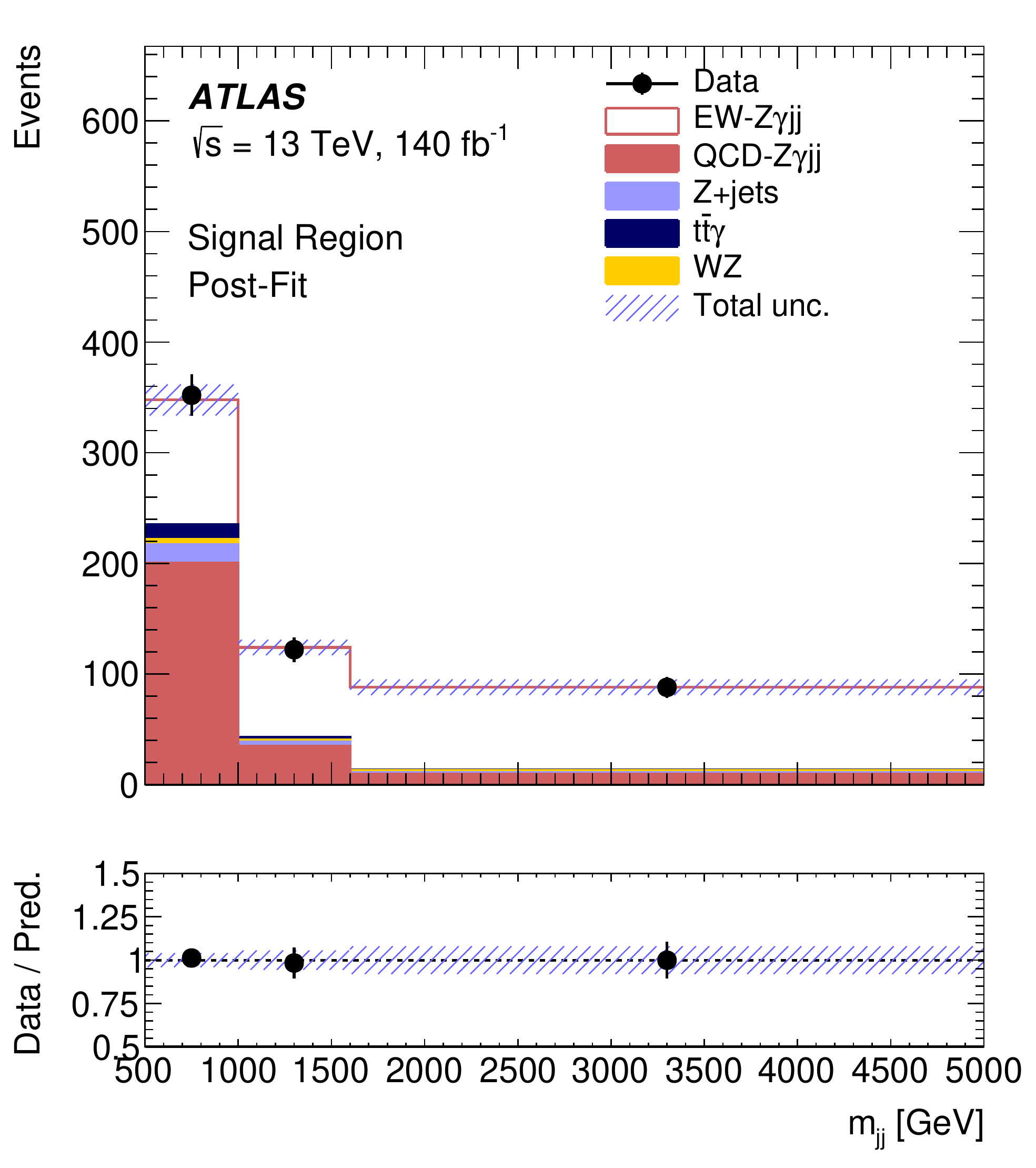}}
\subfloat[]{\includegraphics[width=.35\columnwidth]{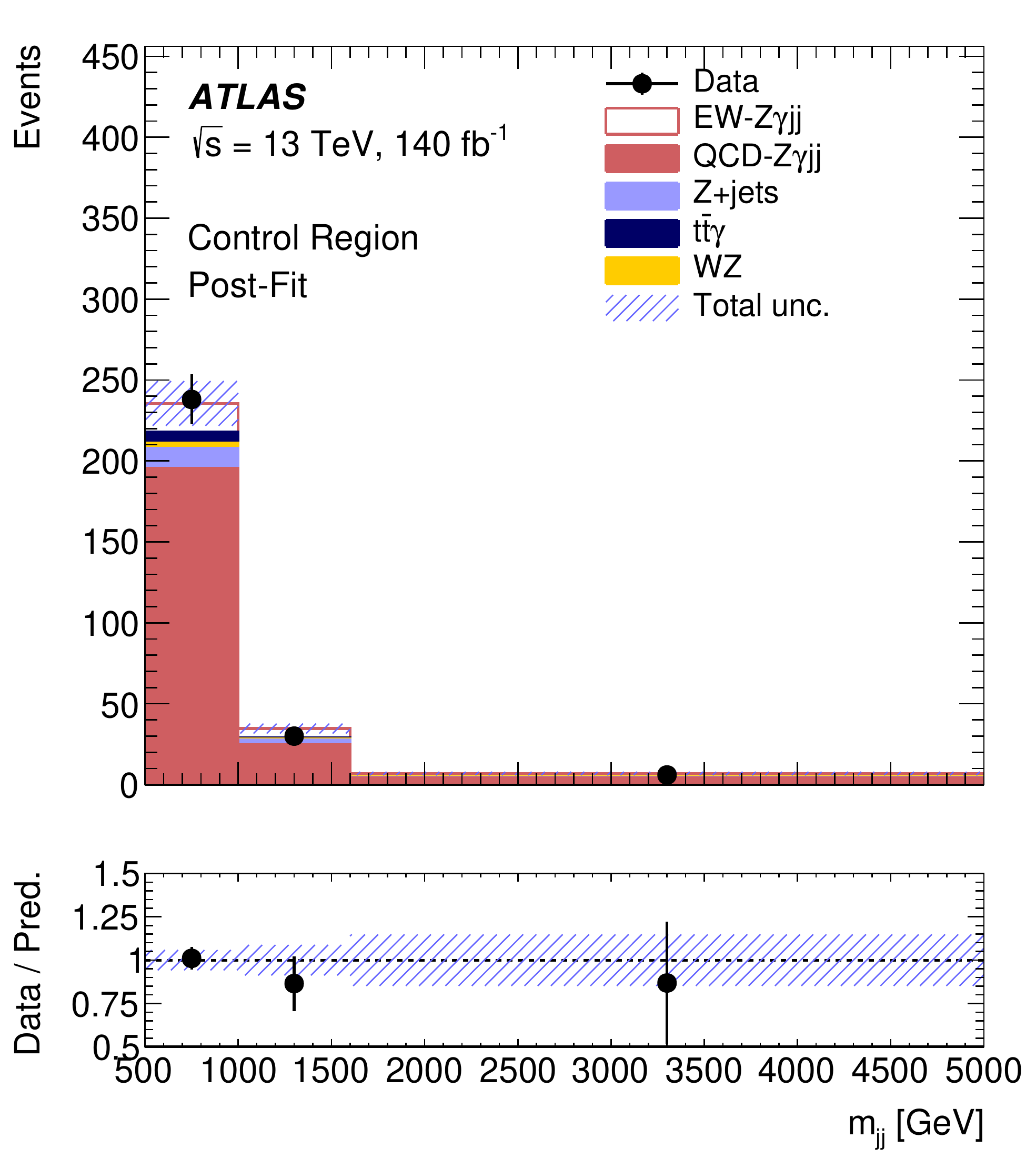}}
\subfloat[]{\includegraphics[width=.35\columnwidth]{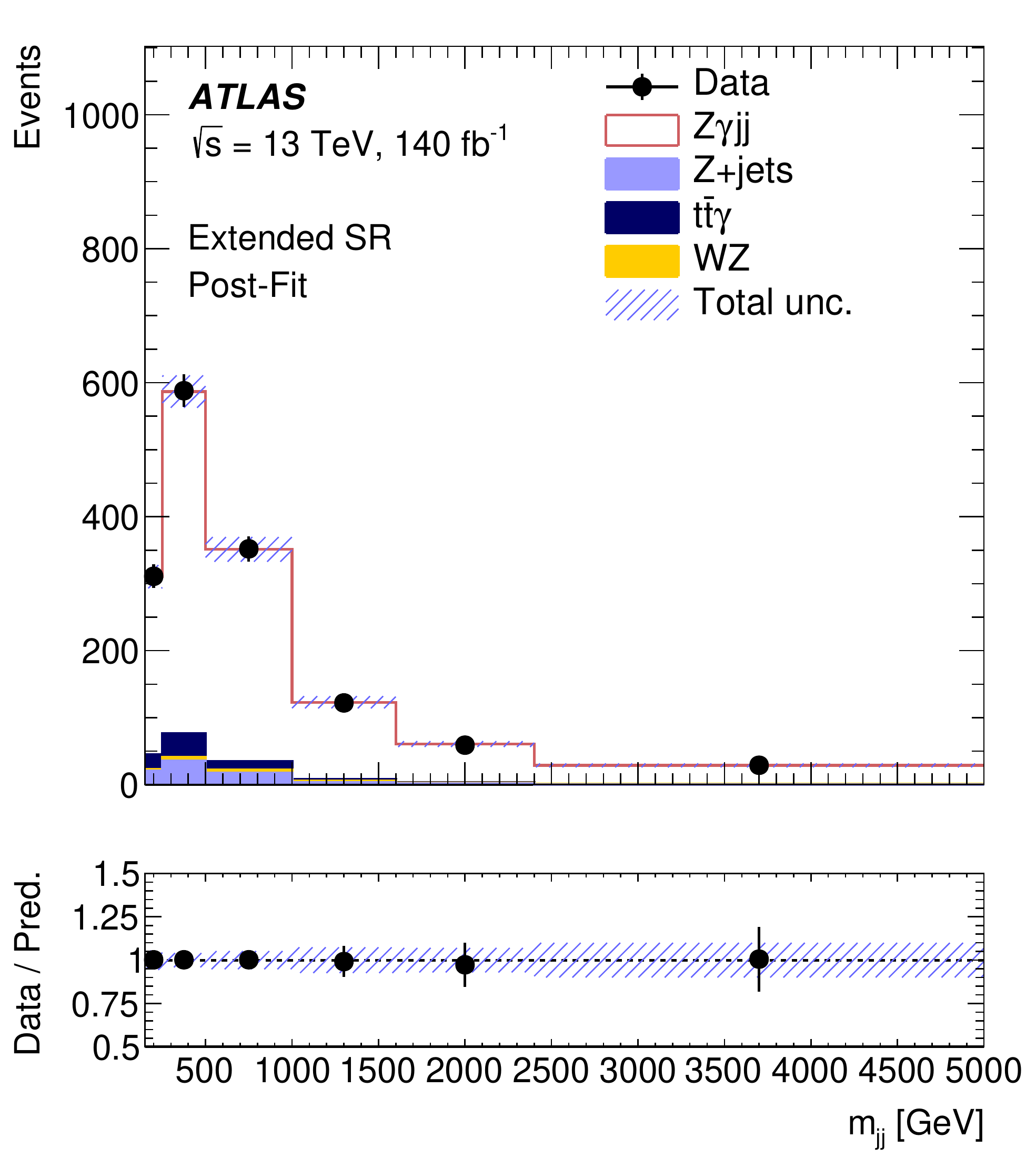}}
\caption{Post-fit $m_{jj}$ distributions in (a) the $m_{jj}>500$~\GeV\ SR (b) the $m_{jj}>500$~\GeV\ CR and (c) the $m_{jj}>150$~\GeV\ Extended SR\@. The uncertainty band around the expectation includes all systematic uncertainties (including MC statistical uncertainty) and takes into account their correlations as obtained from the fit. The error bar around the data points represents the data statistical uncertainty. Events beyond the upper limit of the histogram are included in the last bin.}
\label{fig:postfitPlots}
\end{figure}

Table \ref{tab:BackgroundSummary} shows the observed total number of events and expected number of signal and background events in the SR, Extended SR and CR, after the fit is performed.
The  post-fit $m_{jj}$ distributions are displayed in Figure~\ref{fig:postfitPlots}.

\begin{table}[h!]
\caption{Summary of the observed number of events after the fit in EW signal ($N_{EW-Z\gamma jj}$), QCD $Z\gamma jj$ ($N_{QCD-Z\gamma jj}$), $Z\gamma jj$ ($N_{Z\gamma jj}$), in background ($N_{Z\mathrm{+jets}}$, $N_{t\bar{t}\gamma}$, $N_{WZjj}$), and in data ($N_{\mathrm{obs}}$). The quoted uncertainty corresponds to the post-fit statistical and systematic uncertainties and includes the covariance. The individual uncertainties can be correlated and do not necessarily add in quadrature to equal the total uncertainty.}
\centering
\begin{tabular}{lr@{$\,\pm\,$}lr@{$\,\pm\,$}lr@{$\,\pm\,$}l}
\toprule
Sample & \multicolumn{2}{c}{Ext. SR, $m_{jj}>150$~\GeV} &   \multicolumn{2}{c}{SR, $m_{jj}>500$~\GeV}  & \multicolumn{2}{c}{CR, $m_{jj}>500$~\GeV}\\
\midrule
$N_{\mathrm{EW-Z\gamma jj}}$ & \multicolumn{2}{c}{}  & 269 & 27 & 25 & 6 \\
$N_{\mathrm{QCD-Z\gamma jj}}$ &  \multicolumn{2}{c}{}   & 245 & 21 & 224 & 18 \\
$N_{\mathrm{Z\gamma jj}}$ & 1292 & 50 &  \multicolumn{2}{c}{}  &  \multicolumn{2}{c}{}  \\
$N_{Z\mathrm{+jets}}$          &  78 & 30 & 21 & 8 & 16 & 5  \\
$N_{t\bar{t}\gamma}$    & 73 & 11 & 16 & 2 & 8 & 1  \\
$N_{WZ}$               &  17 & 3 &  9 & 2 & 4 & 1 \\
\midrule
Total                       & 1461 & 38 & 560 & 23 & 277 & 17  \\
\midrule
$N_{\mathrm{obs}}$              &  \multicolumn{2}{l}{1461}  &  \multicolumn{2}{l}{562} &  \multicolumn{2}{l}{274}  \\
 
\bottomrule
\end{tabular}
\label{tab:BackgroundSummary}
\end{table}

% End of text imported from the .//texfiles/8.SignalExtraction.tex input file

% The next lines are included from the .//texfiles/5.PhaseSpace.tex input file
\section{Differential cross-sections}
\label{sec:PhaseSpace}
 
The procedure to extract the $\textit{EW-Z}\gamma jj$  and the $\textit{Z}\gamma jj$  differential cross-sections of the variables discussed in Section~\ref{sec:Introduction} is explained below.
\subsection{Phase space definition}
To define the phase space of the measurement, selection criteria,
which closely mimic the detector-level selection
are applied at particle level to the simulated signal.
This selection
is shown in Table~\ref{tab:fidselection}.

\begin{table}[h!]
\caption{Summary of selection criteria applied at particle level.}
\centering
\begin{tabular}{lc}
\toprule
Lepton  & $p_{\mathrm{T}}^\ell>20, 30 {\mathrm{(leading)}} $~\GeV,  ~~$|\eta_\ell|<2.5$ \\
& $N_{\ell}\geq2$ \\
\midrule
Photon & $E_{\mathrm{T}}^\gamma>25$ \GeV{}, ~~$|\eta_\gamma|<2.37$ \\
& $E_{\mathrm{T}}^{\mathrm{cone20}}<0.07\ET^\gamma$ \\
& $\Delta R(\ell, \gamma)>0.4$ \\
\midrule
Jet  & $p_{\mathrm{T}}^{j}>50$ \GeV{}, ~~$|y_{j}|<4.4$ \\
& $|\Delta y| >1.0$ \\
& $m_{jj}>150$ \GeV{} or $m_{jj}>500$ \GeV{} \\
& Remove jets if $\Delta R(\gamma, j)<0.4$  or if $\Delta R(\ell, j)<0.3$ \\
\midrule
Event    & $m_{\ell \ell} > 40$ \GeV{}  \\
& $m_{\ell \ell}+ m_{\ell \ell \gamma}>182$ \GeV{}  \\
& $\zeta(Z\gamma) < 0.4$  \\
& $N_{\mathrm{jets}}^{\mathrm{gap}} = 0$ \\
\bottomrule
\end{tabular}
 
\label{tab:fidselection}
\end{table}

Only particles with  a mean lifetime $c\tau >10$~mm (referred to as stable particles) are considered.
Only photons and leptons that do not originate from the decay of hadrons  (or, for the leptons, from $\tau$-lepton decays) are selected.
They are referred to as  prompt photons and leptons, respectively.
Contributions from photons within $\Delta R=0.1$ of a lepton are summed together to correct the lepton's four-momentum, a procedure known as `dressing'.
 
At least one isolated photon with transverse momentum $p_{\mathrm{T}} > 25$~\GeV\ and $|\eta|  < 2.37$ is required.
The photon isolation requires that the scalar sum of $p_{\mathrm{T}}$ for all stable particles (except neutrinos, muons and the photon itself) within a cone of radius $\Delta R = 0.2$ around the photon,
$\et^{cone20}$ is  less than 7\% of the photon $p_{\mathrm{T}}$. This  criterion is found to be closest to the detector level isolation criteria  used  in the analysis.
 
The angular distance between  the highest $p_{\mathrm{T}}$ photon and each of the two charged leptons selected  is required to be $\Delta R > 0.4$.
 
Jets are reconstructed using the anti-$k_t$ algorithm with radius parameter $R = 0.4$ using stable particles, excluding neutrinos and prompt electrons, muons and photons.
Jets are considered if their angular distance relative to each of the two charged leptons selected above is  $\Delta R( j, \ell) > 0.3$ and if the angular distance relative to the highest $p_{\mathrm{T}}$  isolated photon is  $\Delta R(j,\gamma)>0.4$.
 
The rejection of events containing $b$-tagged jets as described in Section~\ref{sec:Reconstruction} is not applied in the fiducial phase space. Applying this selection would reduce the predicted $\textit{EW-Z}\gamma jj$ fiducial cross-section by less than  1\% in both SR and CR, and the predicted $\textit{QCD-Z}\gamma jj$ fiducial cross-section event yield by about 7\% in the Extended SR, with no kinematic dependence within the uncertainties of the measurements. The assumption is made that the simulation is correctly extrapolating from reconstructed to fiducial phase space.

\subsection{Unfolding procedure}
 
To obtain the $\textit{EW-Z}\gamma jj$  and the $\textit{Z}\gamma jj$  cross-sections at particle level in the fiducial volume discussed above, an unfolding procedure is performed to correct for detector effects (signal efficiency and acceptance effects).
The unfolding procedure is the same as described in Refs.~\cite{ATLAS:2023xay, ATLAS:2022wec} and based on a profile-likelihood approach.
 
The procedure is applied to the observed number of data events per bin $i$ $N_i^{\mathrm{reco}}$ in the SR and Extended SR, and is related to the number of events at fiducial level in bin $j$ $N_j^{\mathrm{fid}}$ by:
\begin{equation}
\label{eq:Unfold}
N_i^{\mathrm{reco}} = \frac{1}{A_i} \sum_{j} e_{j} M_{ij} N_{j}^{\mathrm{fid}}
\end{equation}
where $M_{ij}$ is the migration matrix
(where each entry represents  the normalised fraction of events at particle level in a bin $j$ that are reconstructed at detector level in a bin $i$ ), $A_{i}$ the acceptance correction (fraction of detector-level  events that are found both in the fiducial volume and in the detector-level selection)
and $e_{j}$ the efficiency correction (fraction of events that are in the fiducial volume that are found both in the fiducial volume and in the detector-level selection). For the $\textit{Z}\gamma jj$ measurement, the migration matrix is built after having summed the EW and QCD contributions.
 
In this procedure, the particle-level bins $j$
are treated as separate subsamples that are multiplied by their respective entries in the response matrix and freely floating parameters ($\mu_{EW}^{j}$ or $\mu_{Z\gamma}^{j}$, the signal strengths defined in Section~\ref{sec:Procedure} applied in bin $j$) are assigned to each of these subsamples at detector level.
In the $\textit{EW-Z}\gamma jj$ unfolding, the CR is fitted simultaneously with the SR to extract the $\textit{QCD-Z}\gamma jj$ bin by bin correction, together with $\mu_{EW}^{j}$.
In this measurement, since the signal contamination is smaller than 1\% in the CR, an approximation is made whereby the signal is treated as an additional background, and no response matrix for the signal is built in the CR\@.
Each bin in the particle-level distribution is  then `folded' through the migration matrix via Eq.~(\ref{eq:Unfold}) to the same number of bins at detector level.
In the unfolding procedure, no regularisation is applied.
 
For the $\textit{EW-Z}\gamma jj$ unfolding, the fraction of events in the diagonal elements of the migration matrix ranges between 80\% ($|\Delta\phi(Z\gamma, jj)|$) and 99\% ($E_T^{\gamma}$, $p_T^{Z\gamma}$, $p_T^{l}$, $|\Delta y|$).
The acceptance corrections are on average around 89\% improving as the variable increases, for all variables except $|\Delta y|$ for which there is no obvious dependence.
The efficiency corrections are at a level of 47\% on average,  and show similar trends as observed for the acceptance corrections.
 
For the $\textit{Z}\gamma jj$ unfolding, the fraction of events in the diagonal elements of the migration matrix ranges between 82\% (|$\Delta \phi(Z\gamma, jj)$|) and 98\% (|$\Delta y$| and $p_T^{\gamma}$).
 
The acceptance corrections are on average around 76\% improving as the variable increases, for all variables except $|\Delta y|$ and $\zeta(Z\gamma)$ for which there is no obvious dependence.
The efficiency corrections are at a level of 40\% on average,  and show similar trends as observed for the acceptance corrections.

The systematic uncertainties considered for the unfolded results are the same as for the results at detector level (see Section~\ref{sec:Systematics}) and are calculated via the migration matrices.
 
Several checks are performed to verify the robustness of the procedure: an injection test with non-SM cross-section values to check if this can be recovered in the unfolding procedure, the use of alternative MC predictions for the $\textit{QCD-Z}\gamma jj$ process and data-driven reweighting of the MC templates using the same observables or alternative ones. None of these checks show any noticeable effect on the unfolding results, and thus no additional uncertainty is assigned to the unfolding procedure.

% End of text imported from the .//texfiles/5.PhaseSpace.tex input file

% The next lines are included from the .//texfiles/9.Results.tex input file
\section{Results}
\label{sec:Results}
The $\textit{EW-Z} \gamma jj$ measured signal strength is
\begin{equation*}
\begin{split}
\mu_{\mathrm{EW}} & = 1.02 \pm 0.09~\text{(stat)} \pm 0.09~\text{(syst)} \\
& =  1.02^{+0.13}_{-0.12}.
\end{split}
\end{equation*}
There is a clear observation of the signal, with a background-only hypothesis rejected with a significance well above 5 standard deviations.
The normalisation parameter of the $\textit{QCD-Z} \gamma jj$ background, constrained by data in the SR and CR is measured to be
$1.18\pm 0.10$.
 
The fiducial  cross-section for the electroweak $pp\rightarrow Z \gamma jj$ process in the phase space defined in Section~\ref{sec:PhaseSpace}
is obtained by computing the product  of the signal strength and the predicted cross-section.
The result is:
\begin{equation*}
\sigma_{\mathrm{EW}} = 3.6 \pm 0.5~{\mathrm{fb}}
\end{equation*}
to be compared with the predicted value from \MGNLO[2.6.5] (interfaced with \PYTHIA) (see Section~\ref{sec:DataMCSamples}), which gives:
\begin{equation*}
\sigma_{\mathrm{EW}}^{pred} = 3.5\pm0.2~\text{fb}.
\end{equation*}
The PDF and scale theoretical uncertainties in the prediction are evaluated using the procedure described in Section~\ref{sec:Systematics}.
 
The breakdown of the systematic uncertainties in the $\textit{EW-Z}\gamma jj$ cross-section is shown in Table~\ref{tab:breakSystData}.
\begin{table}[h!]
\caption{The breakdown of the systematic uncertainties in the $\textit{EW-Z}\gamma jj$ and $Z \gamma jj$ cross-sections. The "Background" component includes uncertainties on $Z+$jets, $t\bar{t}\gamma$ and $WZ$ backgrounds. The "Reco" component includes uncertainties from electrons, photons, muons, jets, flavour tagging and pileup. The "EW mod." component includes interference, parton shower, underlying event, PDF and QCD scale uncertainties in the $\textit{EW-Z}\gamma jj$ process. The "QCD mod." component includes merging scale, resummation scale, PDF and QCD scale uncertainties in the $\textit{QCD-Z}\gamma jj$ process.}
\centering
\begin{tabular}{llllllll}
\toprule
& Data stat.               & MC stat.         & Background    & Reco         & EW mod.          & QCD mod.            & Total \\
\midrule
$\Delta\sigma_{EW}$/$\sigma_{EW}$ [\%]         &  $\pm$9               &       $\pm$1       & $\pm$1          & $\pm$4      &  $^{+8}_{-6}$    &   $\pm 2$       &   $\pm$13   \\
$\Delta\sigma_{Z\gamma}$/$\sigma_{Z\gamma}$ [\%]         &  $\pm$3               &       $\pm$1       & $\pm$2          & $^{+4}_{-3}$      &  $^{+7}_{-6}$    &   $\pm$9        &  $^{+12}_{-11}$    \\
 
\bottomrule
\end{tabular}
\label{tab:breakSystData}
\end{table}

The total cross-section of the process $pp\rightarrow Z \gamma jj$ in the fiducial phase space,
which includes the $\textit{QCD-Z}\gamma jj$ and the  $\textit{EW-Z}\gamma jj$ contributions,
is obtained by multiplying the
signal strength value $\mu_{\mathrm{Z\gamma jj}}$  by the predicted total $Z \gamma jj$ cross-section in the Extended SR\@, where $\mu_{\mathrm{Z\gamma jj}}=1.07 \pm 0.12$.
The measured total $Z \gamma jj$ cross-section is thus:
\begin{equation*}
\sigma_{Z\gamma} = 16.8 ^{+2.0}_{-1.8}~\text{fb},
\end{equation*}
to be compared with the sum of predictions of \MGNLO[2.6.5] interfaced with \PYTHIA (EW contribution) and  \SHERPA 2.2.11 (QCD contribution):
\begin{equation*}
\sigma^{pred}_{Z\gamma} = 15.7 ^{+5.0}_{-2.6}~\text{fb}.
\end{equation*}
 
The PDF and scale theoretical uncertainties in the prediction are evaluated using the procedure described in Section~\ref{sec:Systematics}. Uncertainties are treated as uncorrelated between the $\textit{EW-Z}\gamma jj$ and $\textit{QCD-Z}\gamma jj$ contributions.
 
\begin{figure}[h!]
\centering
\subfloat[]{\includegraphics[width=.4\columnwidth]{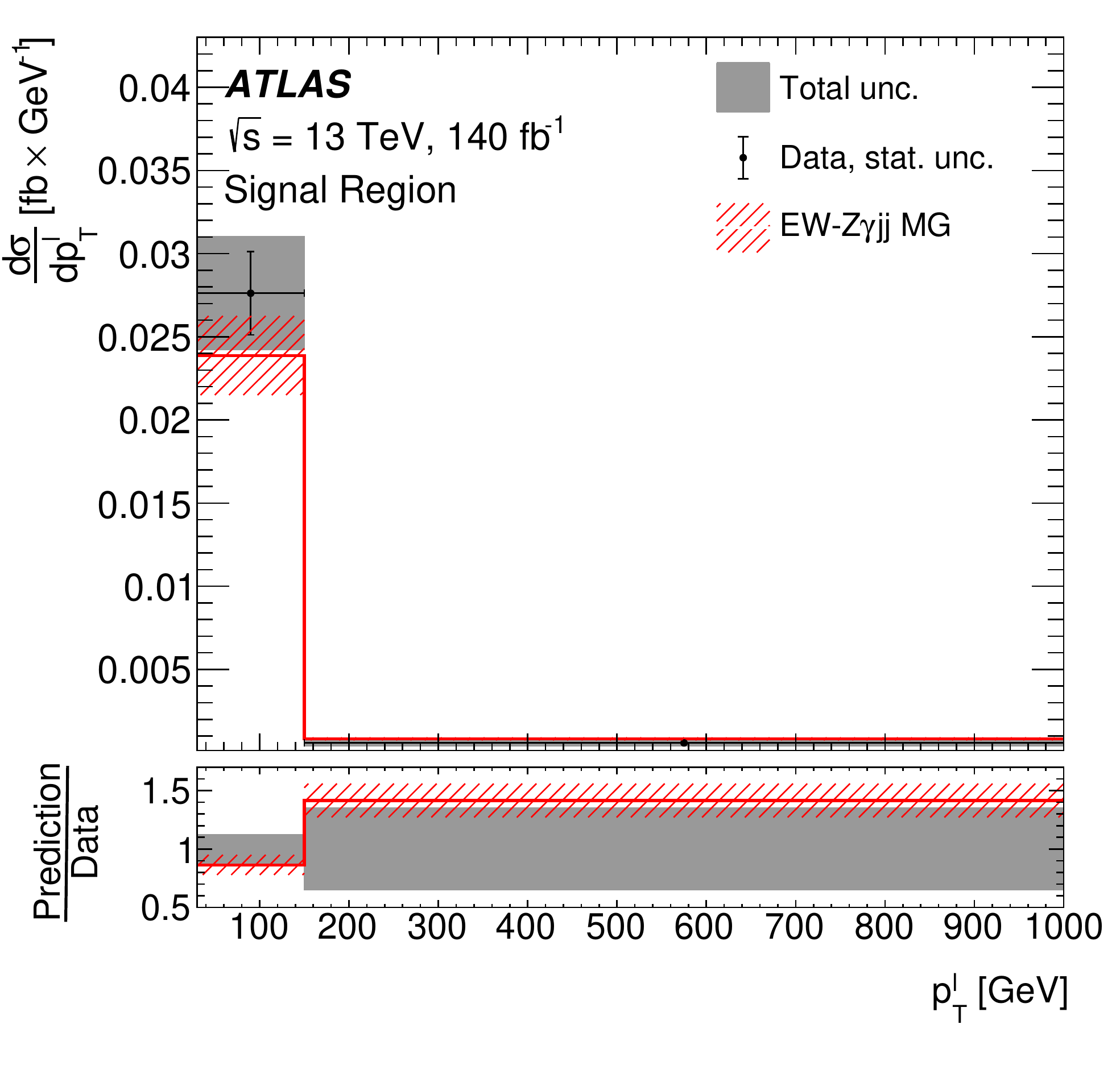}}
\subfloat[]{\includegraphics[width=.4\columnwidth]{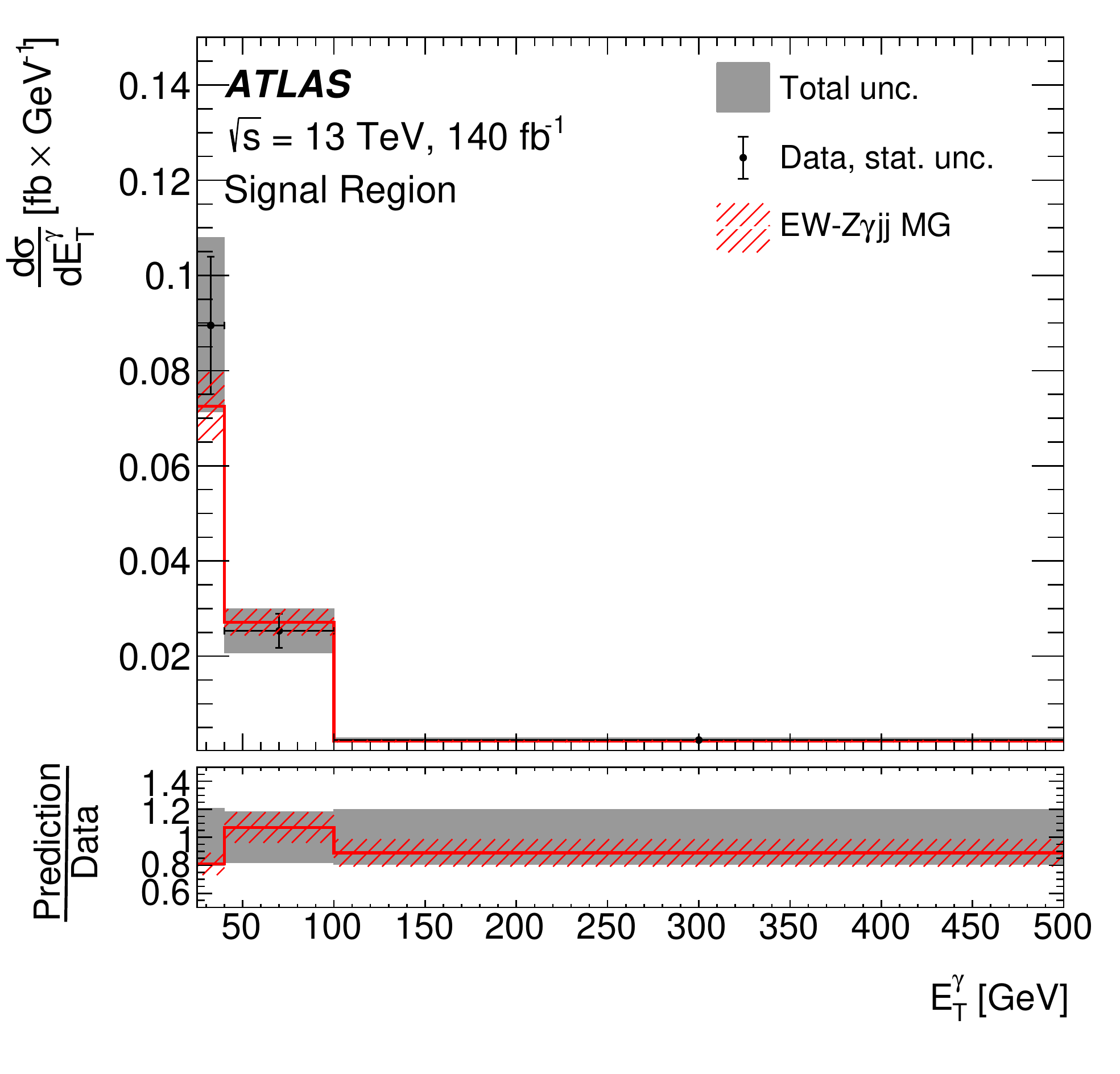}}\\
\subfloat[]{\includegraphics[width=.4\columnwidth]{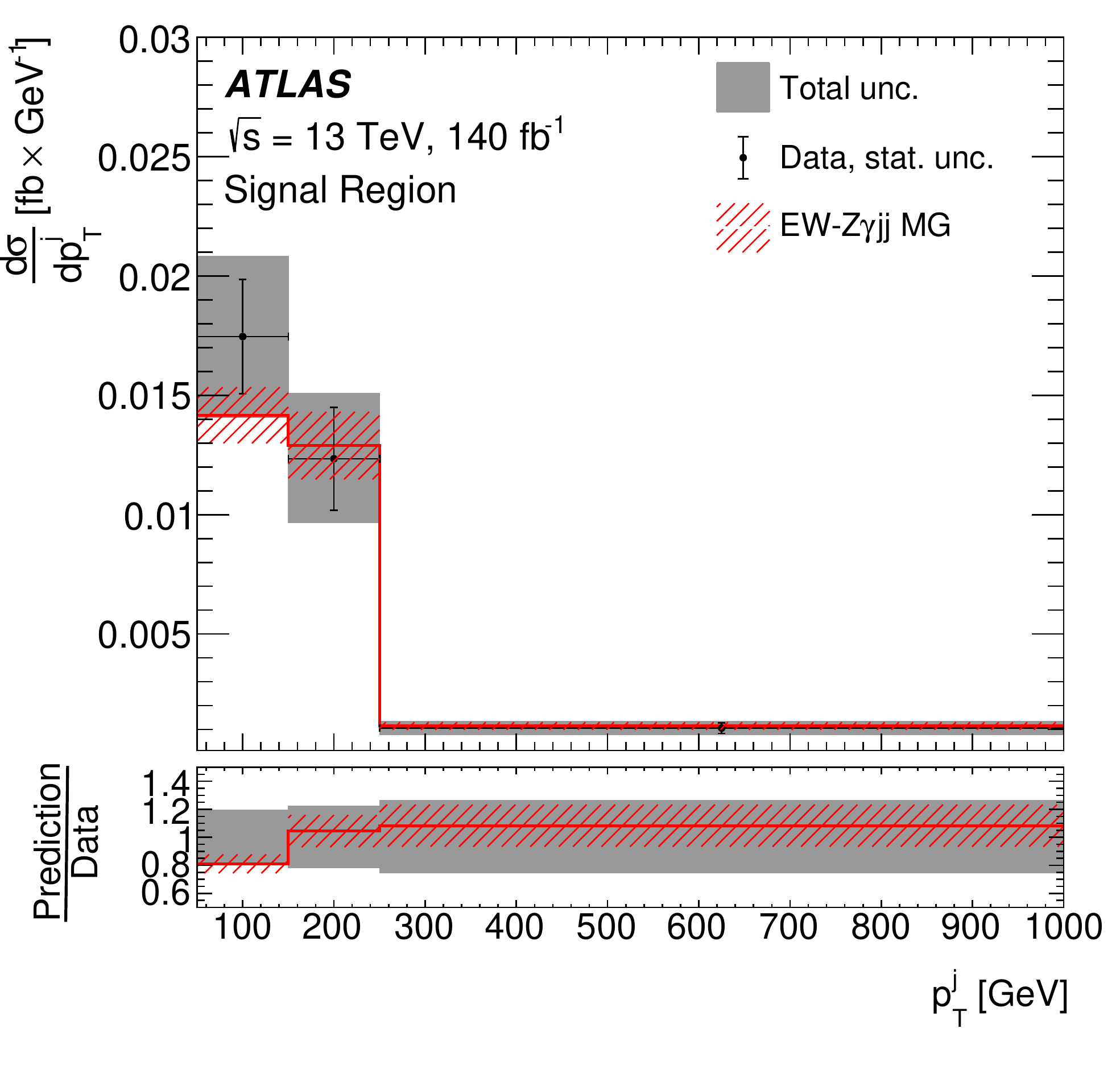}}
\subfloat[]{\includegraphics[width=.4\columnwidth]{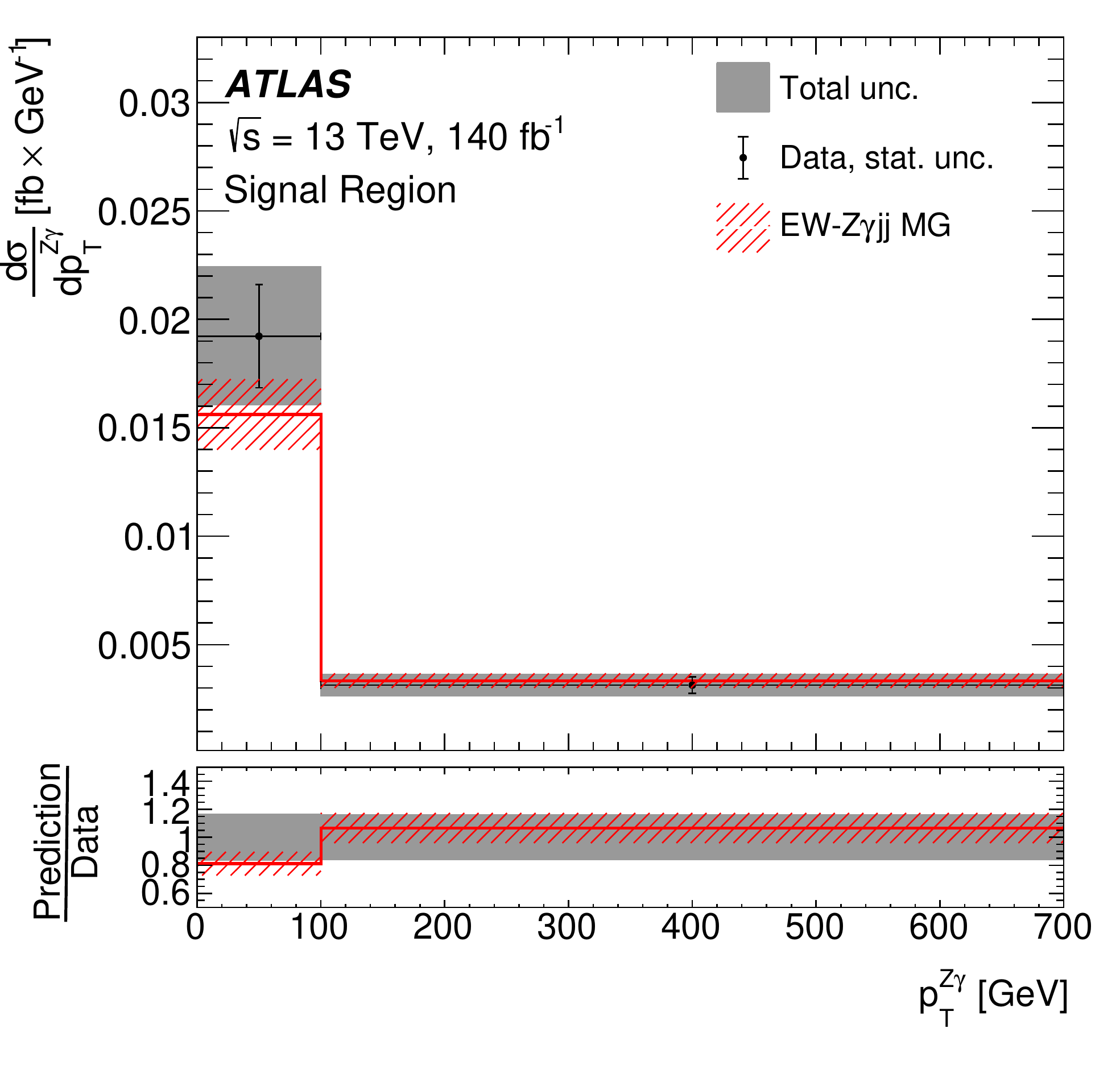}}
\caption{The $\textit{EW-Z}\gamma jj$ differential cross-section in the Signal Region as a function of (a) the leading lepton $p_{T}$, (b) the leading photon $p_{T}$, (c) the leading jet $p_{T}$ and (d) the $Z\gamma$ system $p_{T}$. The lower panels show the ratios of the MC predictions to the data.
The band around the unfolded data represents the total uncertainty (including statistical uncertainty) and takes into account the correlations as obtained from the fit. The hatched area represents the uncertainty in the prediction. Events beyond the upper limit of the histogram are included in the last bin.}
\label{fig:diffXSEW1}
\end{figure}

\begin{figure}[h!]
\centering
\subfloat[]{\includegraphics[width=.4\columnwidth]{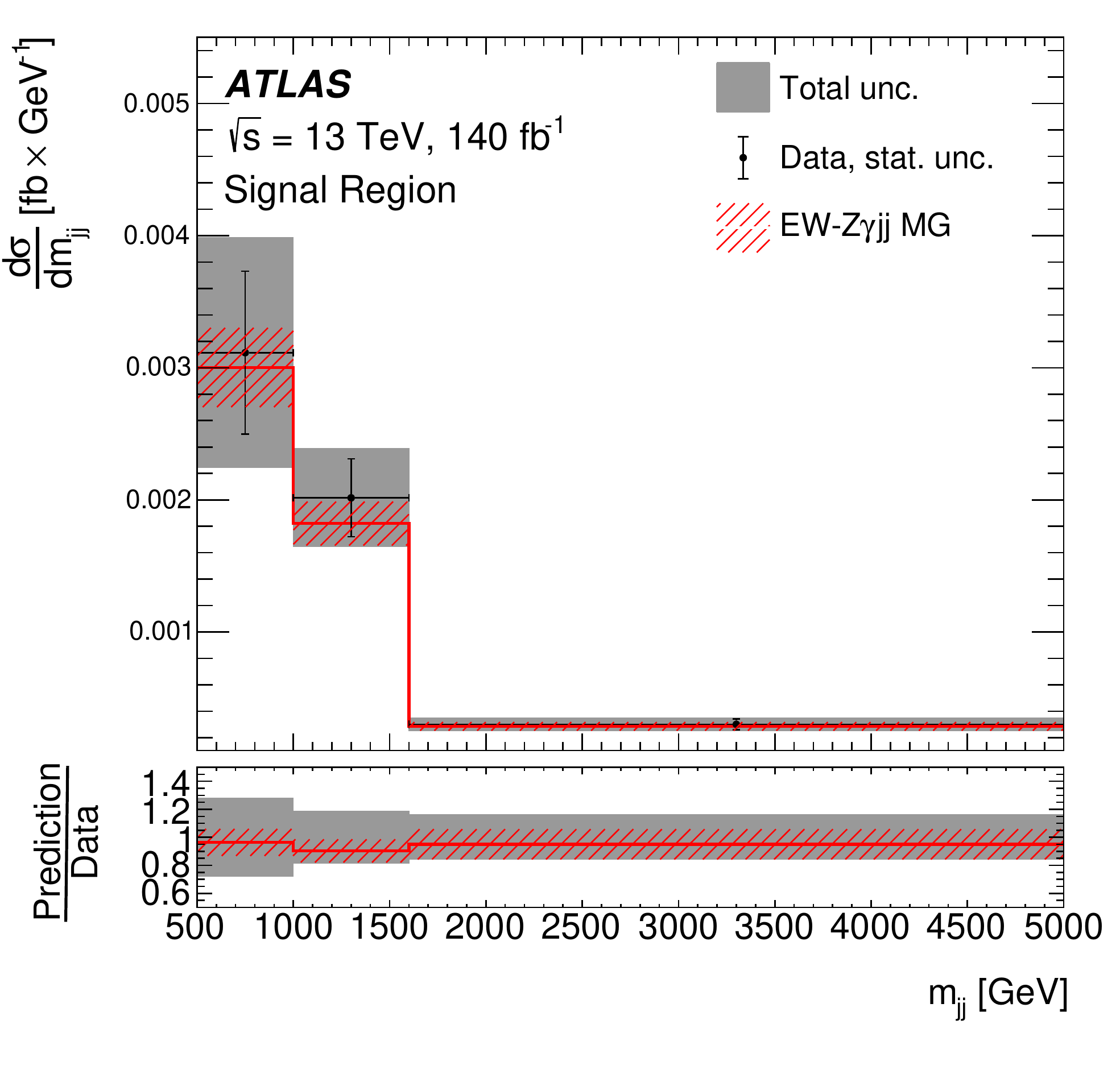}}
\subfloat[]{\includegraphics[width=.4\columnwidth]{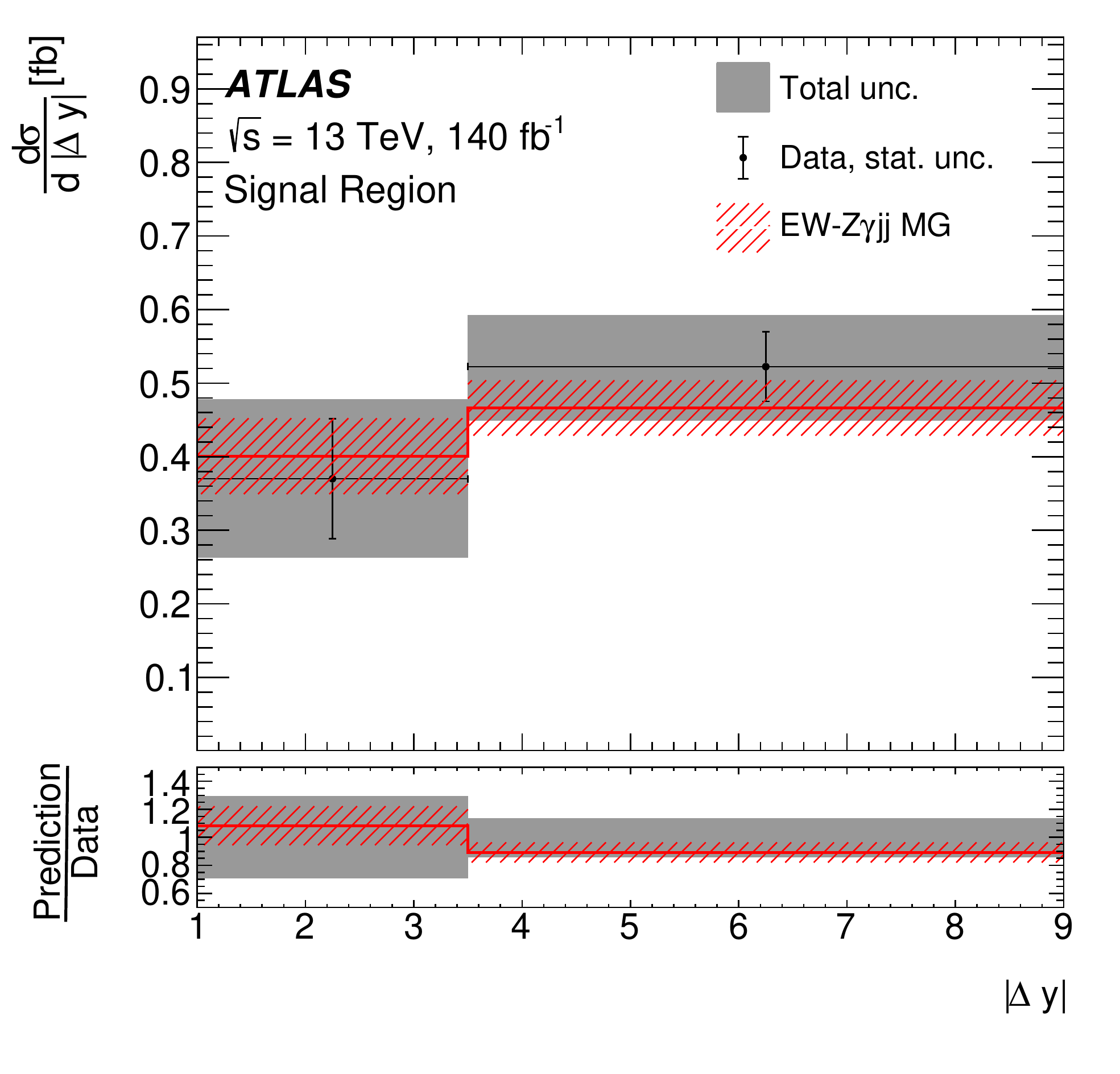}}\\
\subfloat[]{\includegraphics[width=.4\columnwidth]{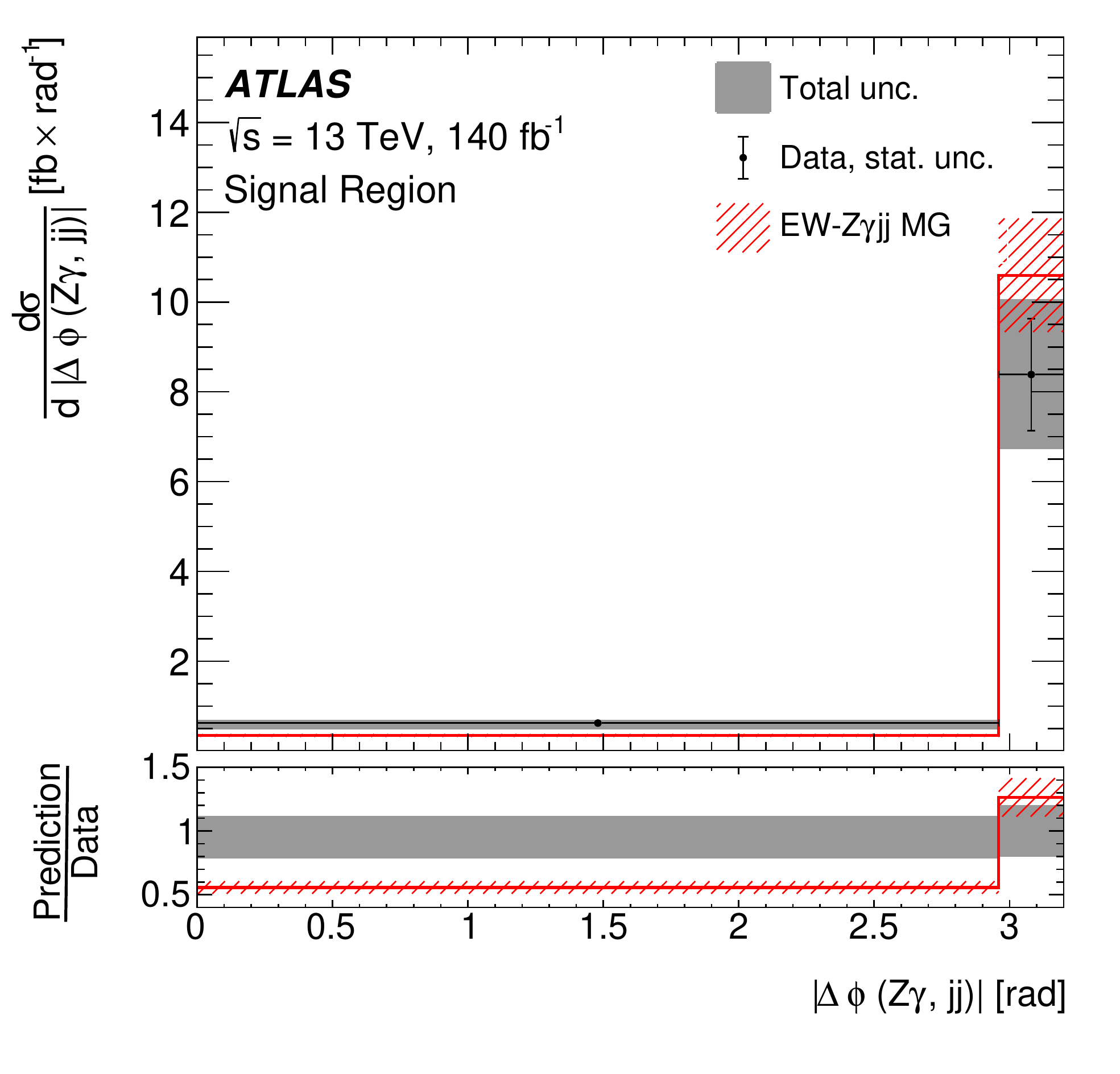}}
\caption{ The $\textit{EW-Z}\gamma jj$ differential cross-section in the Signal Region as a function of (a) the dijet invariant mass, (b) the dijet rapidity difference and (c) the $Z\gamma$ and dijet azimuthal difference. The lower panels show the ratios of the MC predictions to the data.
The band around the unfolded data represents the total uncertainty (including statistical uncertainty) and takes into account the correlations as obtained from the fit. The hatched area represents the uncertainty in the prediction. Events beyond the upper limit of the histogram are included in the last bin.}
\label{fig:diffXSEW2}
\end{figure}

\begin{figure}[h!]
\centering
\subfloat[]{\includegraphics[width=.4\columnwidth]{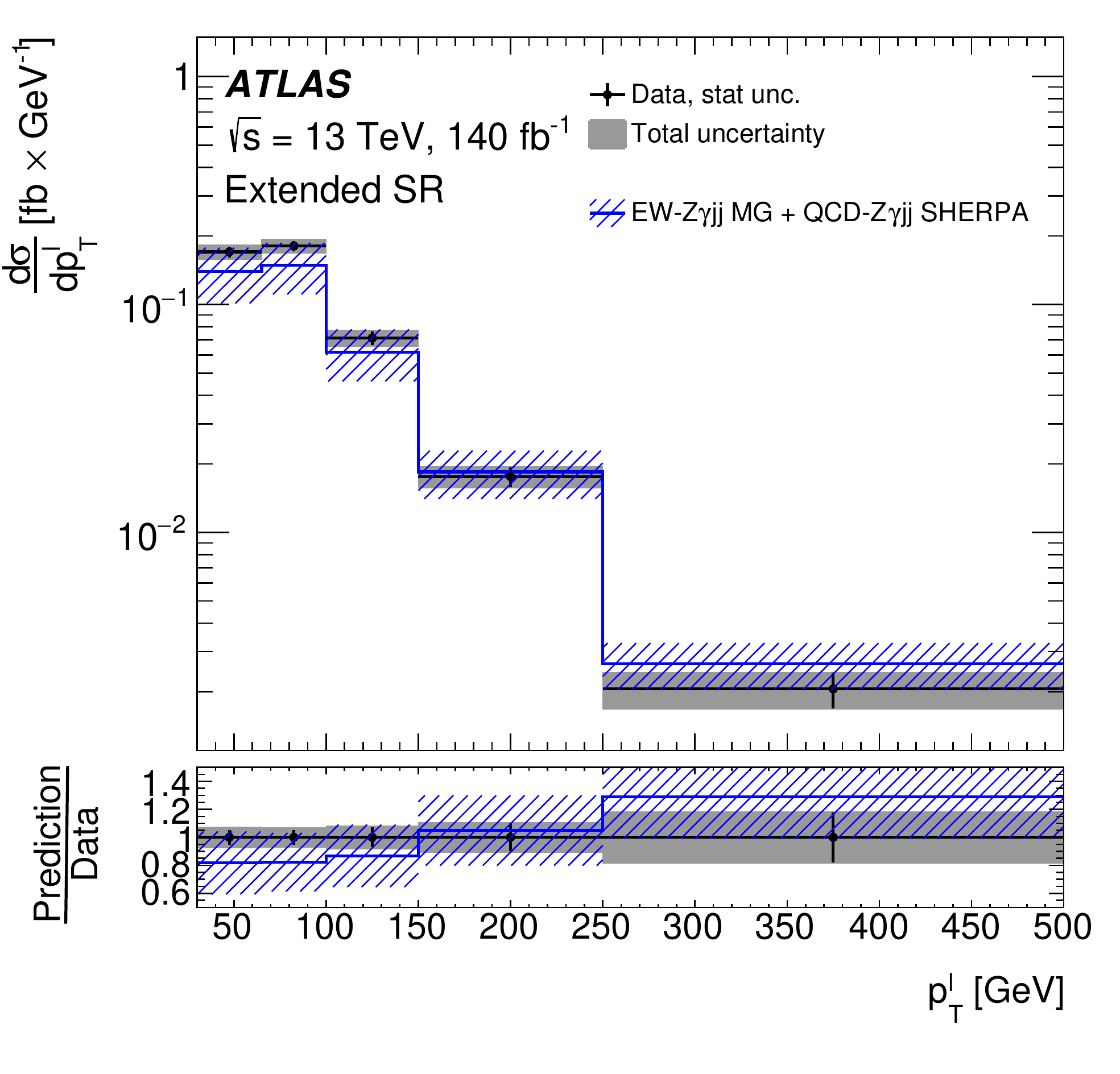}}
\subfloat[]{\includegraphics[width=.4\columnwidth]{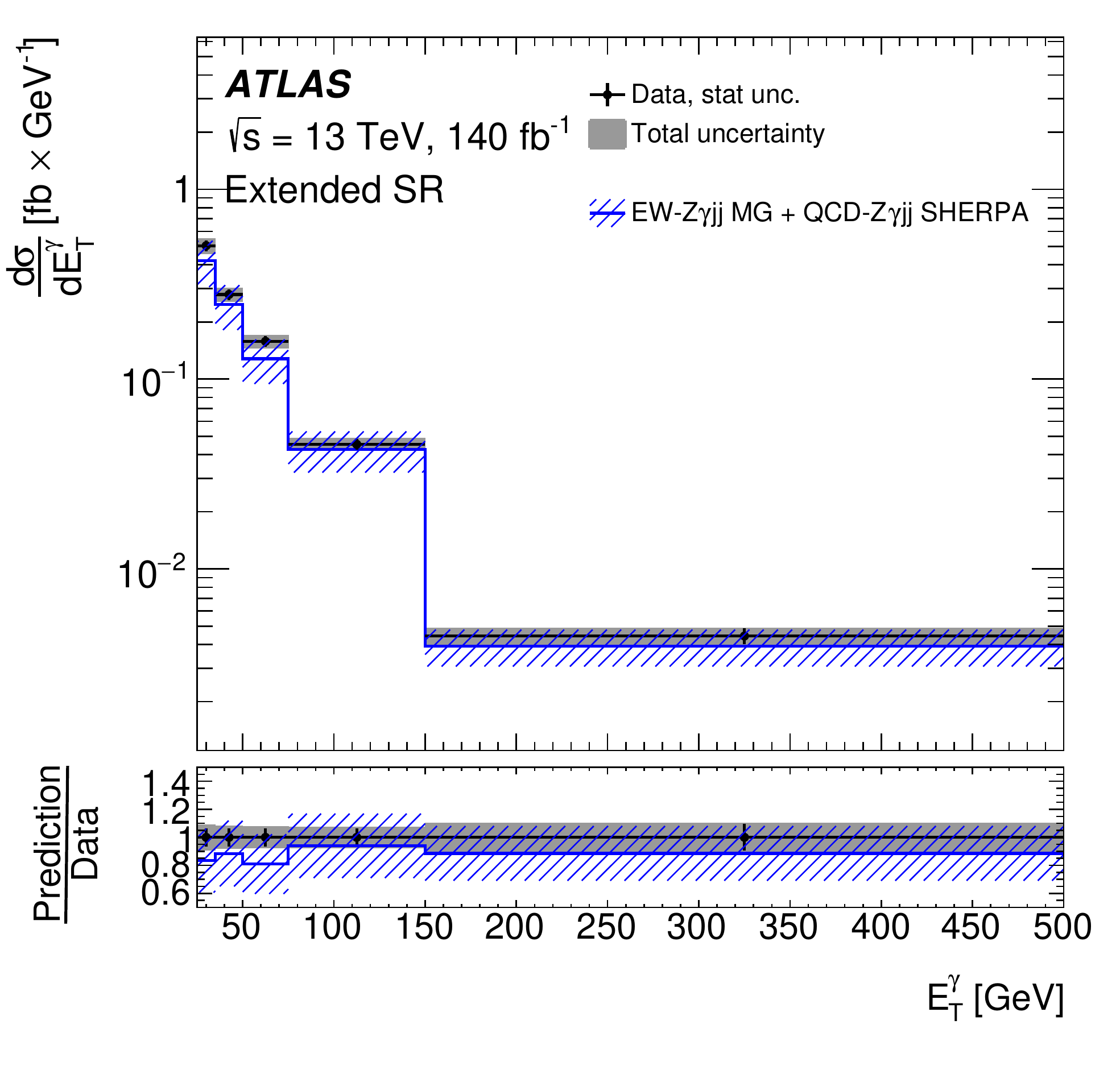}}\\
\subfloat[]{\includegraphics[width=.4\columnwidth]{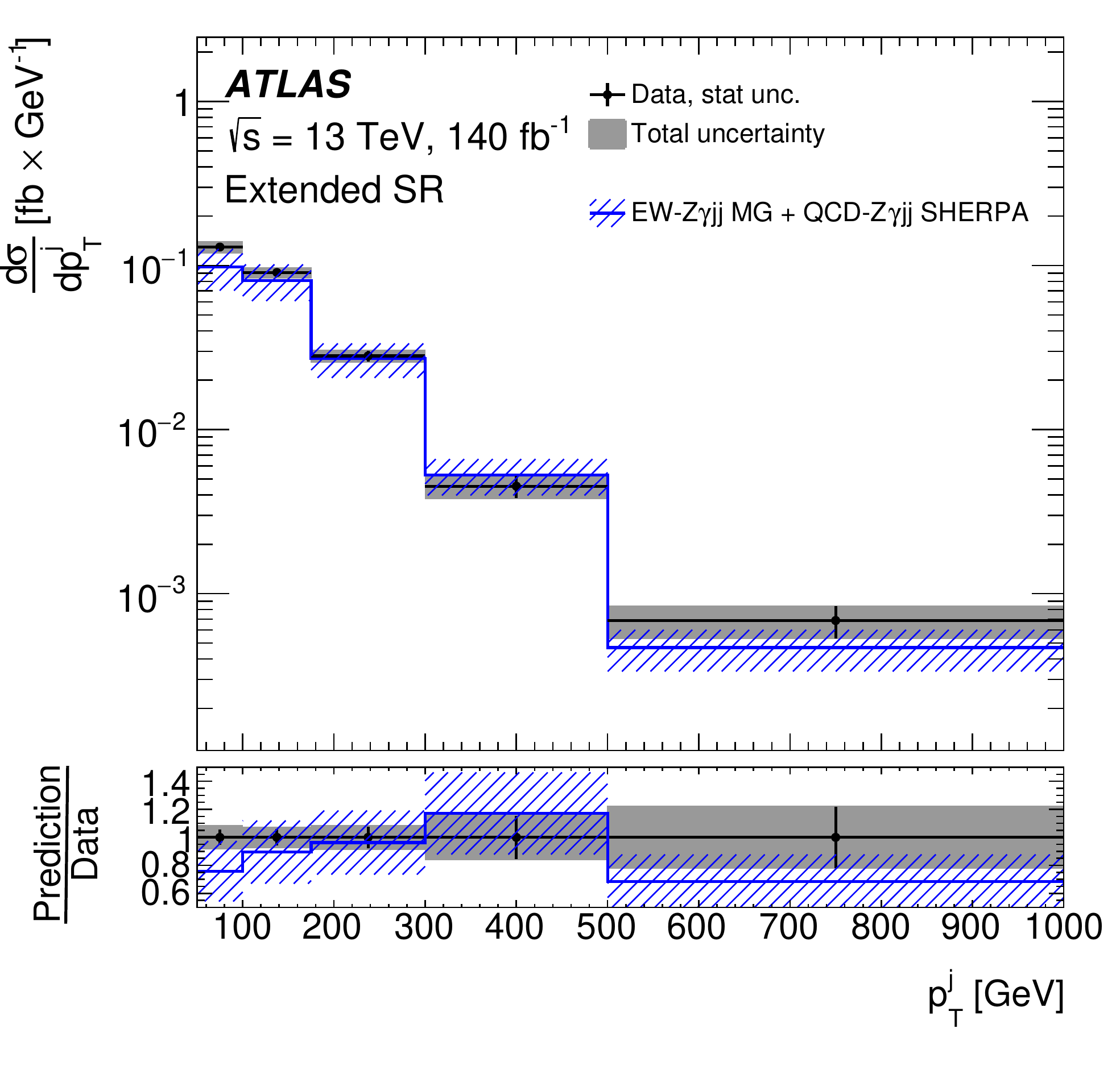}}
\subfloat[]{\includegraphics[width=.4\columnwidth]{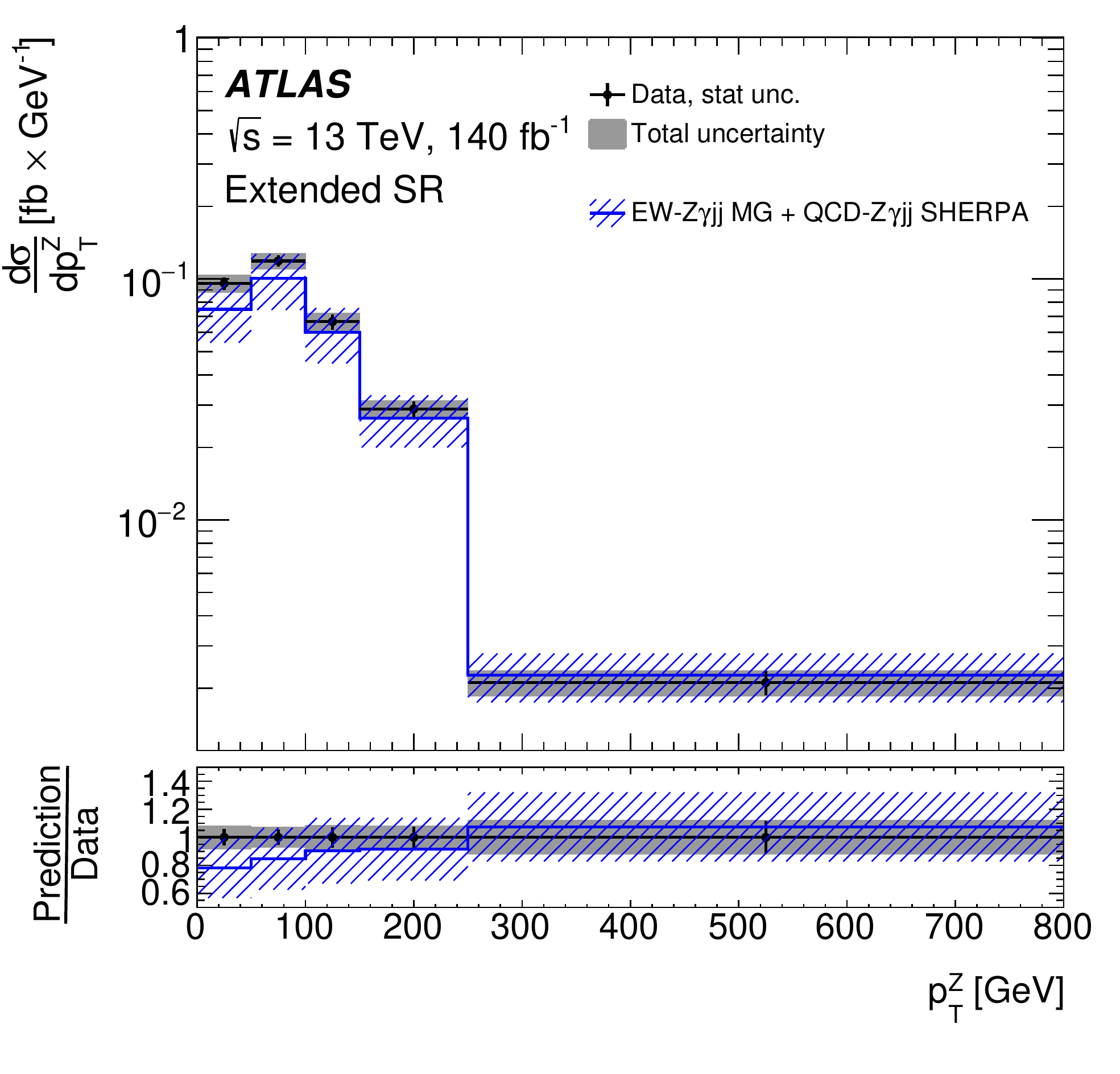}}\\
\subfloat[]{\includegraphics[width=.4\columnwidth]{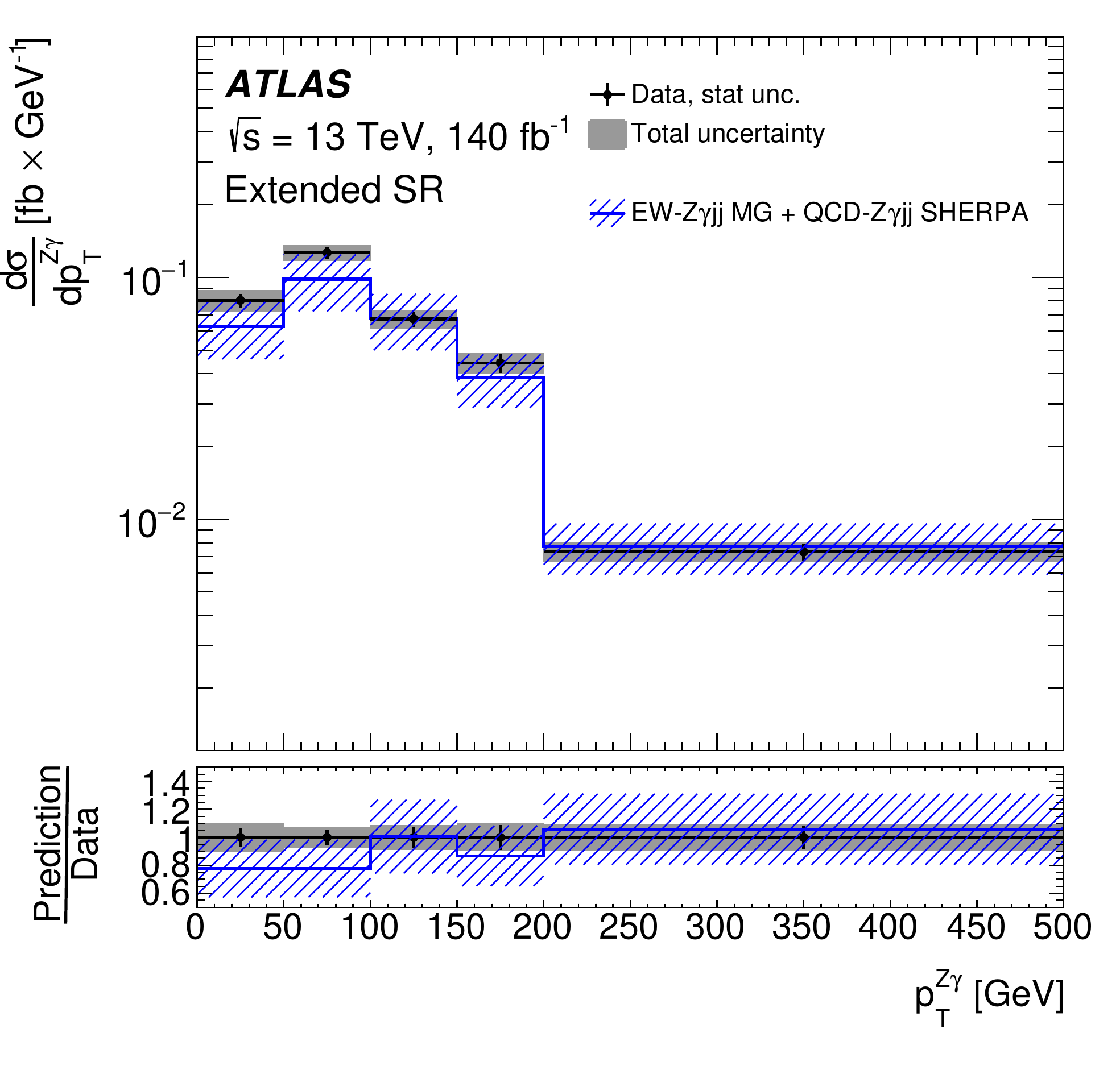}}
\caption{ The total $\textit{Z}\gamma jj$ differential cross-section in the Extended SR as a function of (a) the leading lepton $p_{T}$, (b) the leading photon $p_{T}$, (c) the leading jet $p_{T}$, (d) the $Z$ boson $p_{T}$ and (e) the $Z\gamma$ system $p_{T}$. The lower panels show the ratios of the MC predictions to the data.
The band around the unfolded data represents the total uncertainty (including statistical uncertainty) and takes into account the correlations as obtained from the fit. The hatched area represents the uncertainty in the prediction. Events beyond the upper limit of the histogram are included in the last bin.}
\label{fig:diffXS1}
\end{figure}

\begin{figure}[h!]
\centering
\subfloat[]{\includegraphics[width=.4\columnwidth]{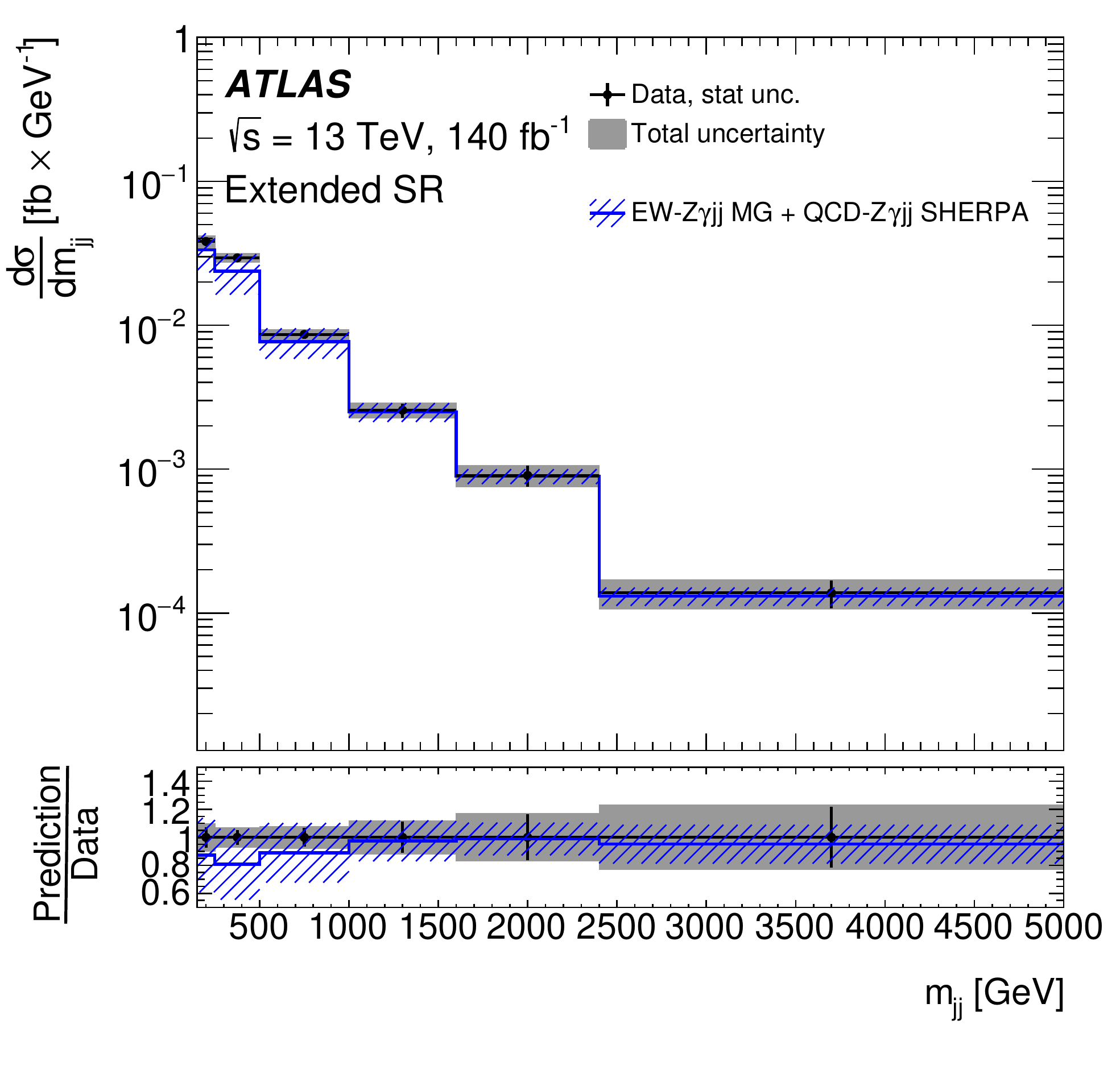}}
\subfloat[]{\includegraphics[width=.4\columnwidth]{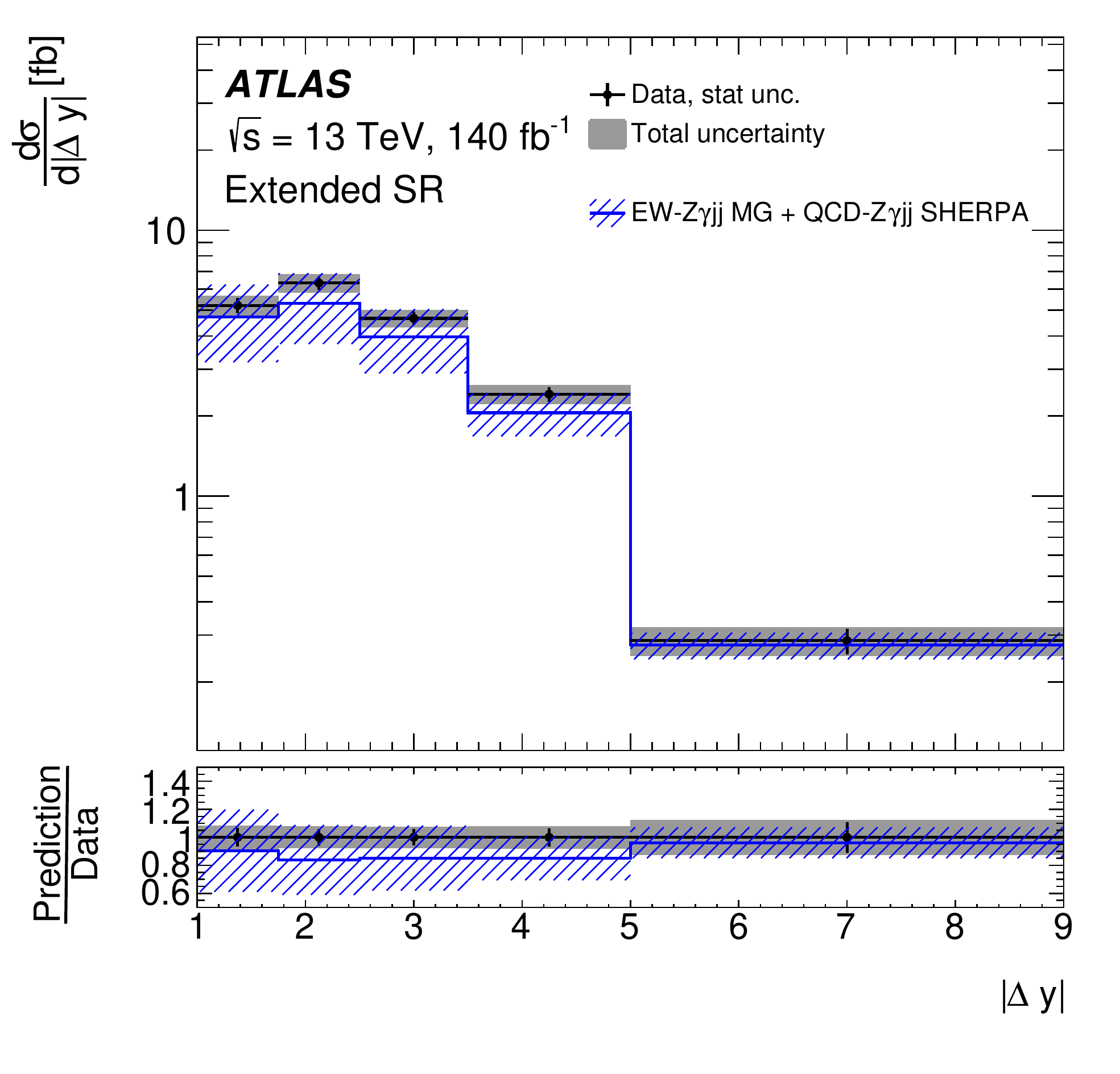}}\\
\subfloat[]{\includegraphics[width=.4\columnwidth]{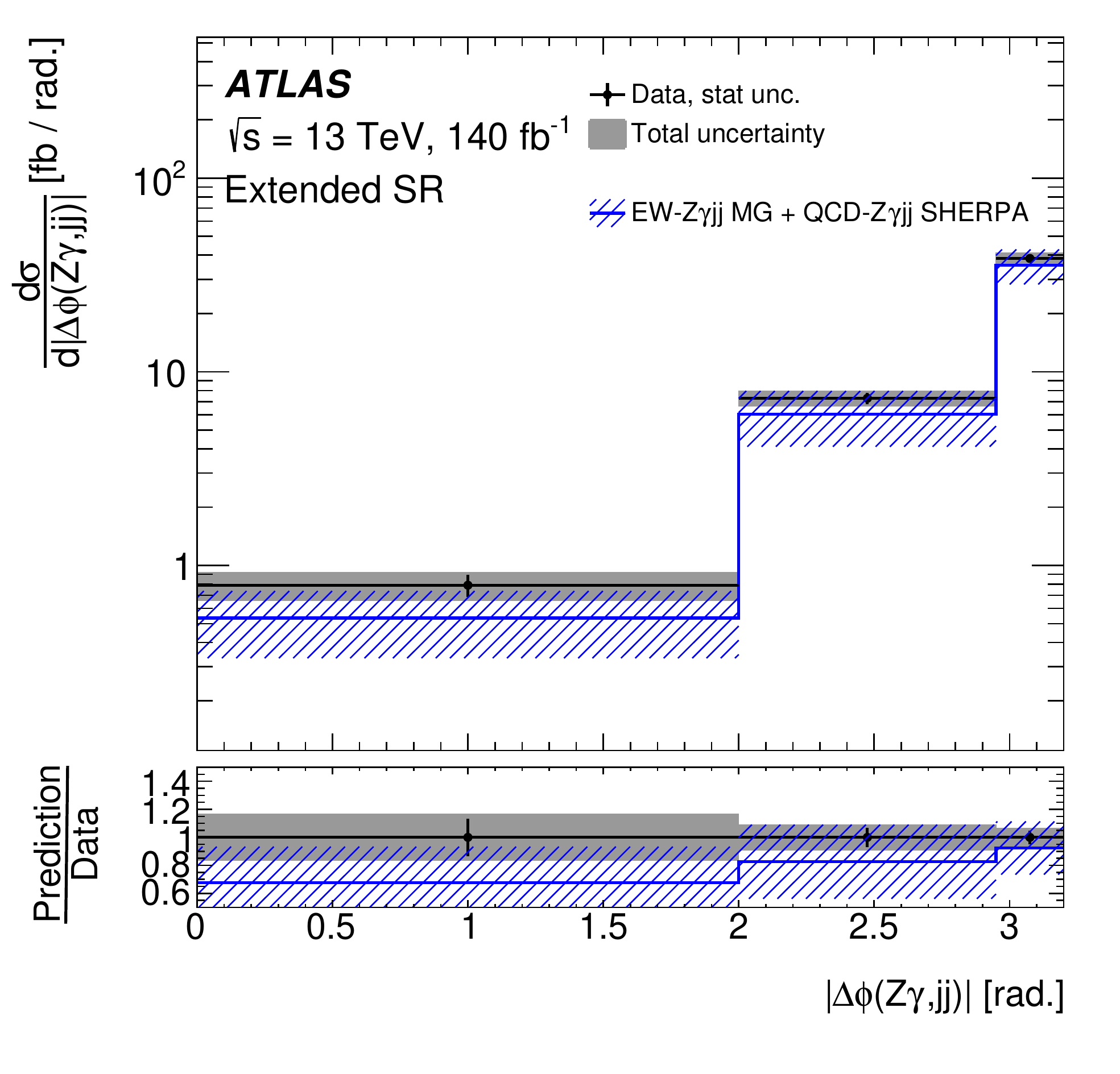}}
\subfloat[]{\includegraphics[width=.4\columnwidth]{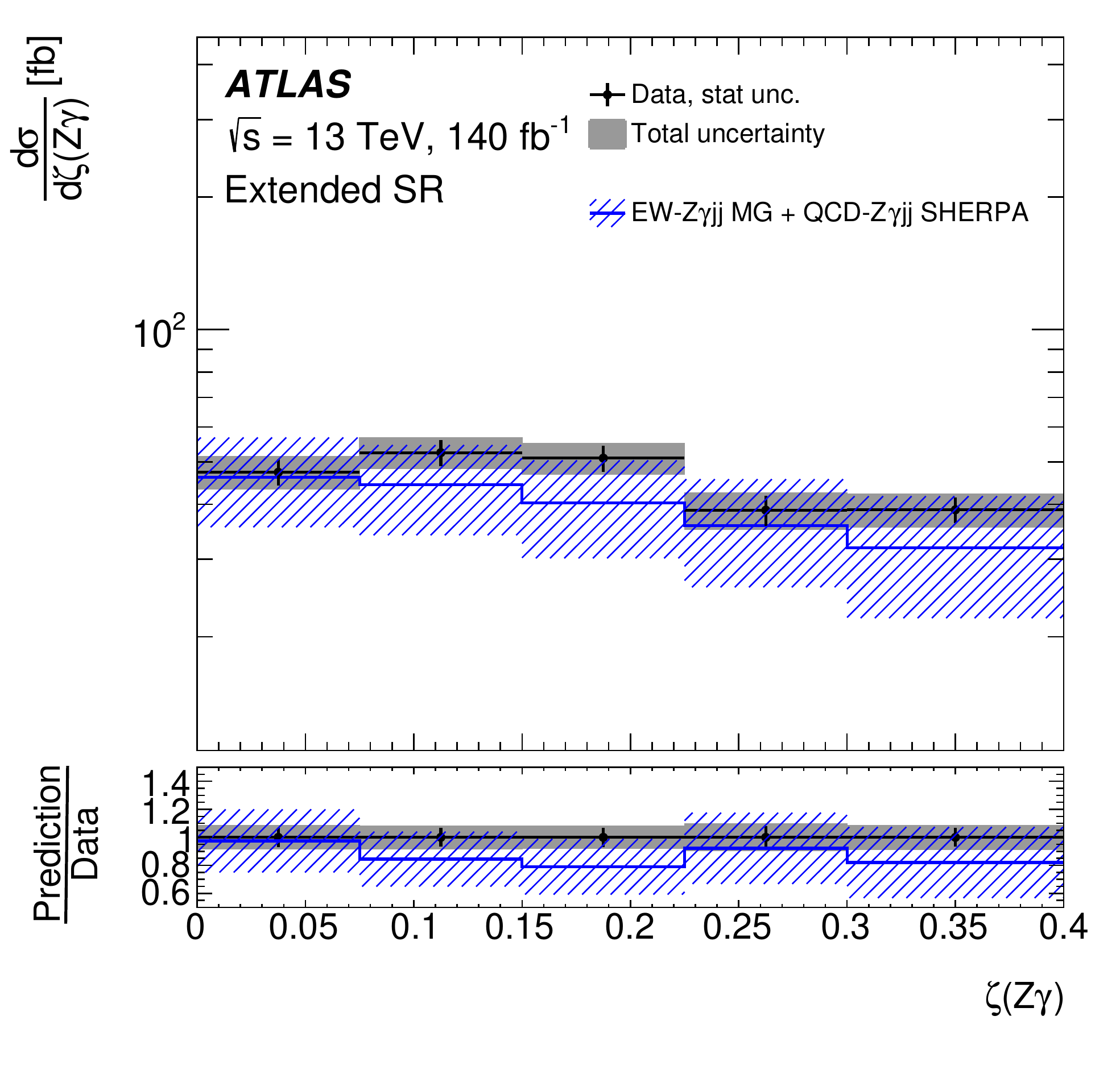}}
\caption{ The total $\textit{Z}\gamma jj$ differential cross-section in the Extended SR as a function of (a) the dijet invariant mass, (b) the dijet rapidity difference, (c) the $Z\gamma$ and dijet azimuthal difference and (d) the centrality of the system $\zeta(Z\gamma)$. The lower panels show the ratios of the MC predictions to the data.
The band around the unfolded data represents the total uncertainty (including statistical uncertainty) and takes into account the correlations as obtained from the fit. The hatched area represents the uncertainty in the prediction. Events beyond the upper limit of the histogram are included in the last bin.}
\label{fig:diffXS2}
\end{figure}

The breakdown of the systematic uncertainties in the $Z \gamma jj$ cross-section is shown in Table~\ref{tab:breakSystData}.
The differential cross-sections are shown in Figures~\ref{fig:diffXSEW1}, ~\ref{fig:diffXSEW2}, ~\ref{fig:diffXS1} and ~\ref{fig:diffXS2}.
 
In the SR phase space, the following variables are measured in two or three bins: $p_{\mathrm{T}}^{l}$, $p_{\mathrm{T}}^{j}$, $E_{\mathrm{T}}^{\gamma}$, $p_{\mathrm{T}}^{Z\gamma}$, $m_{jj}$, $|\Delta y|$ and $|\Delta \phi(Z\gamma, jj)|$. The variables $|\Delta y|$ and $m_{jj}$ are particularly sensitive to the kinematic difference between $\textit{QCD-Z}\gamma jj$ and $\textit{EW-Z}\gamma jj$ events (highest bins being dominated by $\textit{EW-Z}\gamma jj$ events), which make them important variables for VBS studies. They are measured  with a precision ranging from about 25\% (lowest bin) to about 15\% (the highest bin, covering the range 1.5~\TeV\ to 5~\TeV\ for $m_{jj}$ and 3.5 to 9 for $|\Delta y|$).
The variables $p_{\mathrm{T}}^{Z\gamma}$, $E_{\mathrm{T}}^{\gamma}$ and $|\Delta \phi(Z\gamma, jj)|$ are usually studied for their sensitivity to new physics effects~\cite{STDM-2015-21,ATLAS:2022nru,CMS:2021gme}. They are measured in the ranges of 0--700~\GeV, 25--500~\GeV\ and 0--$\pi$, respectively, with a precision of 15--20\% depending on bins and variables. The variables $p_{\mathrm{T}}^{Z\gamma}$ and ($|\Delta \phi(Z\gamma, jj)|$) are in addition measured differentially for the first time at the LHC\@. Transverse momentum $p_{\mathrm{T}}^{l}$ and $p_{\mathrm{T}}^{j}$ are also measured in the ranges of 30--1000~\GeV\ and 50--1000~\GeV\ respectively with a precision of around 25--30\% in the last bin. The \MGNLO predictions reproduce the data well everywhere within uncertainties, except for $|\Delta \phi(Z\gamma, jj)|$ where a $\sim$two standard deviation discrepancy in the lowest bin of  $\textit{EW-Z}\gamma jj$ measurement is seen.
In the Extended SR phase space, the same variables are measured in the same ranges for the total $\textit{Z}\gamma jj$ process with five bins in most cases, except for $m_{jj}$ where the lower range is extended to 150~\GeV, and $p_{T}^{l}$ and $p_{\mathrm{T}}^{Z\gamma}$ for which the higher range is reduced to 500~\GeV.
The measurements in this process are also more precise, on average around 10\%, ranging from $\sim$7\% to $\sim$20\% for lowest to highest $p_{\mathrm{T}}$ bins typically. In addition, $p_{\mathrm{T}}^{Z}$ (from 0 to 800~\GeV) and $\zeta(Z\gamma)$ (from 0 to 0.4) are measured. These two variables are also measured for the first time at the LHC for this process, with a precision of about 10\%. The sum of \MGNLO (for $\textit{EW-Z}\gamma jj$) and  \SHERPA 2.2.11 (for $\textit{QCD-Z}\gamma jj$) predictions reproduce the measurements well within uncertainties.

% End of text imported from the .//texfiles/9.Results.tex input file

\FloatBarrier

% The next lines are included from the .//texfiles/10.Conclusions.tex input file
\section{Conclusions}
\label{sec:Conclusions}

This Letter presents a study of the production of events with a $Z$ boson, decaying into either an $e^+e^-$  or $\mu^+ \mu^-$ pair, a photon and two jets.
The analysis uses 140~fb$^{-1}$ of LHC proton--proton collision data recorded at {\mbox{$\sqrt{s}$ = 13~\GeV}} by the ATLAS detector during the years 2015--2018.
The data sample is enriched in events from the $\textit{EW-Z}\gamma jj$ process by requiring a large dijet invariant mass and by using the information about the centrality of the system. These selections characterise the signal region of the analysis.
The $\textit{EW-Z}\gamma jj$ signal is extracted from a maximum-likelihood fit to the $m_{jj}$ distributions in data  simultaneously using
this signal region and a control region and relying on template MC distributions.
 
The $\textit{EW-Z}\gamma jj$ process is observed by ATLAS in its charged leptonic decay with a significance well above 5~standard deviations by combining the electron and muon channels.
The cross-section of the $\textit{EW-Z}\gamma jj$ process is measured with a precision of 13$\%$ to be {\mbox{3.6$\pm$ 0.5~fb}} in the  signal phase space defined in the analysis, with $m_{jj}>500$~\GeV, to be compared with the predicted value from \MGNLO[2.6.5] which gives {\mbox{3.5$\pm$ 0.3~fb}}.
The $\textit{(EW+QCD)-Z}\gamma jj$ cross-section, which also includes contributions where
the jets arise from the strong interaction, is obtained with a precision of 12$\%$, in the $m_{jj}>150$~\GeV\ phase space.
In the signal phase space of the analysis, the measured $\textit{(EW+QCD)-Z}\gamma jj$ cross-section is {\mbox{16.8$^{+2.0}_{-1.8}$ fb}} to be compared with the sum of predictions of \MGNLO[2.6.5] and  \SHERPA 2.2.11 which gives {\mbox{15.7$^{+5.0}_{-2.6}$ fb}.}
These results are thus consistent with the SM predictions.
Differential cross-section measurements as a function of the transverse momentum of the leading lepton, jet, photon, and $Z\gamma$ system, the invariant mass of and absolute rapidity difference between the two leading jets, the azimuthal difference between $Z\gamma$ system and the two leading jets,  the $Z$ boson transverse momentum, and  the centrality of the $Z\gamma jj$ system are measured for the $\textit{EW-Z}\gamma jj$  and $\textit{Z}\gamma jj$ processes with precision around 20\% and 10\% on average respectively, and all of them are found to be consistent with the SM predictions.

% End of text imported from the .//texfiles/10.Conclusions.tex input file

\section*{Acknowledgements}

% The next lines are included from the .//acknowledgements/Acknowledgements.tex input file

We thank CERN for the very successful operation of the LHC, as well as the
support staff from our institutions without whom ATLAS could not be
operated efficiently.
 
We acknowledge the support of
ANPCyT, Argentina;
YerPhI, Armenia;
ARC, Australia;
BMWFW and FWF, Austria;
ANAS, Azerbaijan;
CNPq and FAPESP, Brazil;
NSERC, NRC and CFI, Canada;
CERN;
ANID, Chile;
CAS, MOST and NSFC, China;
Minciencias, Colombia;
MEYS CR, Czech Republic;
DNRF and DNSRC, Denmark;
IN2P3-CNRS and CEA-DRF/IRFU, France;
SRNSFG, Georgia;
BMBF, HGF and MPG, Germany;
GSRI, Greece;
RGC and Hong Kong SAR, China;
ISF and Benoziyo Center, Israel;
INFN, Italy;
MEXT and JSPS, Japan;
CNRST, Morocco;
NWO, Netherlands;
RCN, Norway;
MEiN, Poland;
FCT, Portugal;
MNE/IFA, Romania;
MESTD, Serbia;
MSSR, Slovakia;
ARRS and MIZ\v{S}, Slovenia;
DSI/NRF, South Africa;
MICINN, Spain;
SRC and Wallenberg Foundation, Sweden;
SERI, SNSF and Cantons of Bern and Geneva, Switzerland;
MOST, Taiwan;
TENMAK, T\"urkiye;
STFC, United Kingdom;
DOE and NSF, United States of America.
In addition, individual groups and members have received support from
BCKDF, CANARIE, Compute Canada and CRC, Canada;
PRIMUS 21/SCI/017 and UNCE SCI/013, Czech Republic;
COST, ERC, ERDF, Horizon 2020 and Marie Sk{\l}odowska-Curie Actions, European Union;
Investissements d'Avenir Labex, Investissements d'Avenir Idex and ANR, France;
DFG and AvH Foundation, Germany;
Herakleitos, Thales and Aristeia programmes co-financed by EU-ESF and the Greek NSRF, Greece;
BSF-NSF and MINERVA, Israel;
Norwegian Financial Mechanism 2014-2021, Norway;
NCN and NAWA, Poland;
La Caixa Banking Foundation, CERCA Programme Generalitat de Catalunya and PROMETEO and GenT Programmes Generalitat Valenciana, Spain;
G\"{o}ran Gustafssons Stiftelse, Sweden;
The Royal Society and Leverhulme Trust, United Kingdom.
 
The crucial computing support from all WLCG partners is acknowledged gratefully, in particular from CERN, the ATLAS Tier-1 facilities at TRIUMF (Canada), NDGF (Denmark, Norway, Sweden), CC-IN2P3 (France), KIT/GridKA (Germany), INFN-CNAF (Italy), NL-T1 (Netherlands), PIC (Spain), ASGC (Taiwan), RAL (UK) and BNL (USA), the Tier-2 facilities worldwide and large non-WLCG resource providers. Major contributors of computing resources are listed in Ref.~\cite{ATL-SOFT-PUB-2023-001}.

% End of text imported from the .//acknowledgements/Acknowledgements.tex input file

\clearpage
\printbibliography
 
\clearpage
% ATLAS Collaboration author list
% Reference date of STDM-2018-36 is 2023-04-28
% Author list last updated on date 05-DEC-23
% Data extracted on 05-Dec-2023 for paper reference STDM-2018-36
% at 6:45pm
 
\begin{flushleft}
\hypersetup{urlcolor=black}
{\Large The ATLAS Collaboration}

\bigskip

\AtlasOrcid[0000-0002-6665-4934]{G.~Aad}$^\textrm{\scriptsize 102}$,
\AtlasOrcid[0000-0002-5888-2734]{B.~Abbott}$^\textrm{\scriptsize 120}$,
\AtlasOrcid[0000-0002-1002-1652]{K.~Abeling}$^\textrm{\scriptsize 55}$,
\AtlasOrcid[0000-0001-5763-2760]{N.J.~Abicht}$^\textrm{\scriptsize 49}$,
\AtlasOrcid[0000-0002-8496-9294]{S.H.~Abidi}$^\textrm{\scriptsize 29}$,
\AtlasOrcid[0000-0002-9987-2292]{A.~Aboulhorma}$^\textrm{\scriptsize 35e}$,
\AtlasOrcid[0000-0001-5329-6640]{H.~Abramowicz}$^\textrm{\scriptsize 151}$,
\AtlasOrcid[0000-0002-1599-2896]{H.~Abreu}$^\textrm{\scriptsize 150}$,
\AtlasOrcid[0000-0003-0403-3697]{Y.~Abulaiti}$^\textrm{\scriptsize 117}$,
\AtlasOrcid[0000-0003-0762-7204]{A.C.~Abusleme~Hoffman}$^\textrm{\scriptsize 137a}$
% ...
\bigskip
\\

% ...
$^{29}$Physics Department, Brookhaven National Laboratory, Upton NY; United States of America.\\
% ...
$^{35}$$^{(a)}$Facult\'e des Sciences Ain Chock, R\'eseau Universitaire de Physique des Hautes Energies - Universit\'e Hassan II, Casablanca;$^{(b)}$Facult\'{e} des Sciences, Universit\'{e} Ibn-Tofail, K\'{e}nitra;$^{(c)}$Facult\'e des Sciences Semlalia, Universit\'e Cadi Ayyad, LPHEA-Marrakech;$^{(d)}$LPMR, Facult\'e des Sciences, Universit\'e Mohamed Premier, Oujda;$^{(e)}$Facult\'e des sciences, Universit\'e Mohammed V, Rabat;$^{(f)}$Institute of Applied Physics, Mohammed VI Polytechnic University, Ben Guerir; Morocco.\\
% ...
$^{49}$Fakult\"{a}t Physik , Technische Universit{\"a}t Dortmund, Dortmund; Germany.\\
% ...
$^{55}$II. Physikalisches Institut, Georg-August-Universit\"{a}t G\"ottingen, G\"ottingen; Germany.\\
% ...
$^{102}$CPPM, Aix-Marseille Universit\'e, CNRS/IN2P3, Marseille; France.\\
% ...
$^{117}$Department of Physics, New York University, New York NY; United States of America.\\
% ...
$^{120}$Homer L. Dodge Department of Physics and Astronomy, University of Oklahoma, Norman OK; United States of America.\\
% ...
$^{137}$$^{(a)}$Departamento de F\'isica, Pontificia Universidad Cat\'olica de Chile, Santiago;$^{(b)}$Millennium Institute for Subatomic physics at high energy frontier (SAPHIR), Santiago;$^{(c)}$Instituto de Investigaci\'on Multidisciplinario en Ciencia y Tecnolog\'ia, y Departamento de F\'isica, Universidad de La Serena;$^{(d)}$Universidad Andres Bello, Department of Physics, Santiago;$^{(e)}$Instituto de Alta Investigaci\'on, Universidad de Tarapac\'a, Arica;$^{(f)}$Departamento de F\'isica, Universidad T\'ecnica Federico Santa Mar\'ia, Valpara\'iso; Chile.\\
% ...
$^{150}$Department of Physics, Technion, Israel Institute of Technology, Haifa; Israel.\\
$^{151}$Raymond and Beverly Sackler School of Physics and Astronomy, Tel Aviv University, Tel Aviv; Israel.\\
% ...
$^{a}$ Also Affiliated with an institute covered by a cooperation agreement with CERN.\\
$^{b}$ Also at An-Najah National University, Nablus; Palestine.\\
$^{c}$ Also at Borough of Manhattan Community College, City University of New York, New York NY; United States of America.\\
$^{d}$ Also at Center for High Energy Physics, Peking University; China.\\
$^{e}$ Also at Center for Interdisciplinary Research and Innovation (CIRI-AUTH), Thessaloniki; Greece.\\
$^{f}$ Also at Centro Studi e Ricerche Enrico Fermi; Italy.\\
% ...

\end{flushleft}

% Created with Glance <Atlas.Glance@cern.ch>

\end{document}